%% file: sample-acmsmall.tex
\documentclass[acmsmall, manuscript, screen]{acmart}
\usepackage{bm}
\usepackage{multirow}
\usepackage{ulem}
\usepackage{dsfont}
\usepackage{bbm}
\usepackage{enumitem}

\newtheorem{problem}{Problem}

\AtBeginDocument{%
  \providecommand\BibTeX{{%
    \normalfont B\kern-0.5em{\scshape i\kern-0.25em b}\kern-0.8em\TeX}}}

\setcopyright{acmcopyright}
\copyrightyear{2018}
\acmYear{2018}
\acmDOI{XXXXXXX.XXXXXXX}

\acmJournal{JACM}
\acmVolume{37}
\acmNumber{4}
\acmArticle{111}
\acmMonth{8}

\begin{document}
\title{Heterogeneous Information Crossing on Graphs for Session-based Recommender Systems}

\author{Xiaolin Zheng}
\affiliation{%
\institution{College of Computer Science, Zhejiang University}
\city{Hangzhou}
\country{China}
}
\email{xlzheng@zju.edu.cn}

\author{Rui Wu}
\affiliation{%
\institution{College of Computer Science, Zhejiang University}
\city{Hangzhou}
\country{China}
}
\email{CS\_wurui@163.com}

\author{Zhongxuan Han}
\affiliation{%
\institution{College of Computer Science, Zhejiang University}
\city{Hangzhou}
\country{China}
}
\email{zxhan@zju.edu.cn}

\author{Chaochao Chen*}\thanks{*Corresponding author. }
\affiliation{%
  \institution{College of Computer Science, Zhejiang University}
  \city{Hangzhou}
  \country{China}
}
\email{zjuccc@zju.edu.cn}

\author{Linxun Chen}
\affiliation{%
\institution{MYbank, Ant Group}
\city{Hangzhou}
\country{China}
}
\email{linxun.clx@antgroup.com}

\author{Bing Han}
\affiliation{%
\institution{MYbank, Ant Group}
\city{Hangzhou}
\country{China}
}
\email{hanbing.hanbing@alibaba-inc}



\input{./content/0_abstract.tex}


\maketitle
\newpage

\input{./content/content.tex}

\normalem
\bibliographystyle{ACM-Reference-Format}
\bibliography{sample-base}

\end{document}

%% file: content/0_abstract.tex
\begin{abstract}
Recommender systems are fundamental information filtering techniques 
to recommend content or items that meet users' personalities and potential needs.
As a crucial solution to address the difficulty of user identification and unavailability of historical information, session-based recommender systems provide recommendation services that only rely on users' behaviors 
in the current session.
However, most existing studies are not well-designed for modeling heterogeneous user behaviors and capturing the relationships between them in practical scenarios. 
To fill this gap, in this paper, we propose a novel graph-based method, 
namely \textbf{H}eterogeneous \textbf{I}nformation \textbf{C}rossing on \textbf{G}raphs (HICG).
HICG utilizes multiple types of user behaviors in the sessions to construct heterogeneous graphs, 
and captures users' current interests with their long-term preferences by effectively crossing the heterogeneous information on the graphs.
In addition, we also propose an enhanced version, named HICG-CL, which incorporates contrastive learning (CL) technique to enhance item representation ability.
By utilizing the item co-occurrence relationships across different sessions, HICG-CL improves the recommendation performance of HICG.
We conduct extensive experiments on three real-world recommendation datasets, and the results verify that (i) HICG 
achieves the state-of-the-art performance by utilizing multiple types of behaviors on the heterogeneous graph.
(ii) HICG-CL further significantly improves the recommendation performance of HICG by the proposed contrastive learning module. 
\end{abstract}

%% file: content/content.tex
\input{./content/1_introduction/content.tex}
\input{./content/2_related_work/content.tex}
\input{./content/3_method/content.tex}
\input{./content/4_experiments/content.tex}

\input{./content/5_conclusion.tex}
\input{./content/6_acknowledgments.tex}

%% file: content/1_introduction/content.tex
\section{Introduction} \label{ss:1}

\begin{figure}[htbp]
    \centering
    \includegraphics[width=1\linewidth,height=0.3\textwidth]{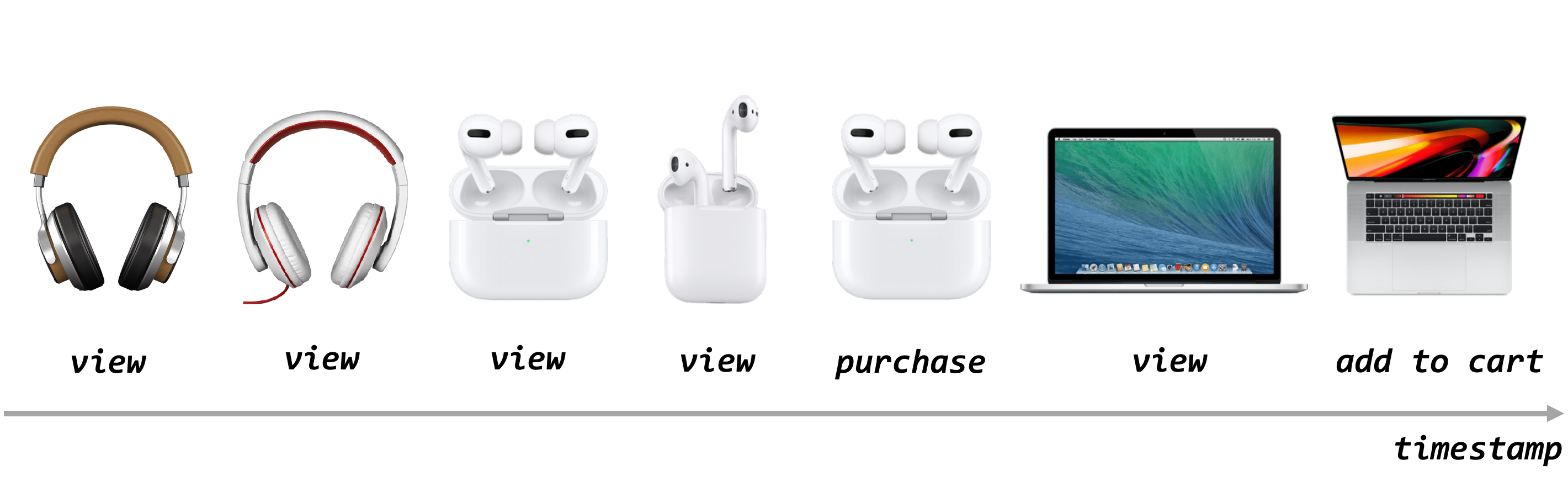}
    \caption{An example session of user sequential behaviors in the e-commerce scenario.}
    \label{fig:sample}
\end{figure}

Nowadays, recommender systems (RS) play a critical role in many real-life services such as e-commerce, streaming platforms, and social networks \cite{zhu2021unified}.
Conventional recommender systems usually assume that long-term user profiles and user-item historical interactions are available ~\cite{Cao2015CouplingLO,Wang2020TheEO,Lops2011ContentbasedRS,zhu2021cross}.
However, in many practical scenarios, user identification and historical information may not be available (except those in the current session) due to certain reasons, e.g., privacy regulation ~\cite{Quadrana2017PersonalizingSR, Ren2019RepeatNetAR}.
Recently, session-based recommender systems (SBRS) are proposed to overcome this problem, which utilizes the limited historical behaviors in each ongoing session to learn customer preferences and provide recommendations~\cite{Wang2019ASO, Dang2020ASO}.

Due to the highly practical value, researchers have proposed kinds of session-based recommendation methods in the past few years. 
Most existing methods employ Markov chains ~\cite{Shani2002AnMR,Rendle2010FactorizingPM,Zimdars2001UsingTD} or recurrent neural networks (RNN)  ~\cite{Hidasi2016SessionbasedRW,Tan2016ImprovedRN,narm2017jing} to infer user preferences from the transitional patterns in their interacted items.
First, Markov chains based models assume that a user's subsequent behavior depends on the items he interacted in the latest behavior.
However, it assumes that independence only exists between subsequent behaviors and thence neglects the high-order relationships within the behavior sequence, which limits the recommendation performance. 
Secondly, to relieve this assumption, researchers have proposed RNNs-based models to capture  the sequential properties, which leads to significant progress. 
Among them, GRU4Rec ~\cite{Hidasi2016SessionbasedRW} is the first work in session-based recommendation that applies gated recurrent unit (GRU) to capture transitional patterns in user behavior sequence. 
By introducing an attention mechanism to capture the user's primary purpose in the current session, NARM \cite{narm2017jing} further improves the effects of GRU4Rec.
STAMP ~\cite{Liu2018STAMPSA} also achieves good performance by applying an attention mechanism with simple multi-layer networks to capture users' potential preferences and current interests.
Recently, researchers have utilized graph neural networks (GNN) to achieve state-of-the-art performance across multiple fields ~\cite{Wu2019ACS}. 
There are also attempts, such as SR-GNN~\cite{Wu2019SessionbasedRW}, GC-SAN~\cite{Xu2019GraphCS}, and others~\cite{Xia2021SelfSupervisedHC,Yu2021SelfSupervisedMH,peng2021lime,peng2021reinforced}, apply GNNs to improve the recommendation performance by modeling the complex transition relationships between items. 

Although the above mentioned methods have made significant progress on SBRS, they focus on modeling a single type of user interaction and fail to incorporate the multiple types.
In reality, SRBS can collect various types of interactions, such as browsing, adding-to-cart, purchasing, dwelling time, and ratings.
In fact, various types of behaviors are essential for SRBS to model a user's preference in the current session, and ignoring this will inevitably result in estimation bias. 
\autoref{fig:sample} shows an example session in e-commerce scenario where the current user has three types of behaviors, including view, add-to-cart, and purchase.
Although this user may require earphones (or headphones), as he browsed extensively at the beginning, the purchase of AirPods and his subsequent behaviors on the MacBook products indicate that his need for earphones has been satisfied.
In short, there are two things that should be noticed by SBRS: 1. The act of purchasing represents the current user's satisfaction with the need for headphones; 2. The subsequent behaviors of laptops indicate the user's new interests or demands, which should be taken care of by SBRS, and the importance of those behaviors should be different from previous ones in the following recommendations.
Ignoring multiple types of user behaviors will fail to capture a user's interest drift behind this signal and finally encounter poor performance when repeatedly recommending the same type of items, e.g., earphones in this example.
In recent years, a few SRBS work ~\cite{Meng2020IncorporatingUM,Wang2019ModelingMS,Wang2020BeyondCM} has been proposed to utilize multi-behavior in given sessions.
However, those methods use heterogeneous information by either combining individual behavior prediction tasks or simply concatenating the representations that are extracted from different types of behaviors.
Without considering the underlying relationships between heterogeneous behaviors, these methods still have limited performance.

%
To adequately model the complex sequential relations between users' interacted items and capture the relationships among different types of behaviors, we propose a novel session-based recommendation method called \textbf{H}eterogeneous \textbf{I}nformation \textbf{C}rossing on \textbf{G}raphs (\textbf{HICG}).
HICG integrates various types of user behaviors on heterogeneous graphs and explores users' current interests with their preferences on these graphs simultaneously.
Specifically, for each session, HICG first builds a heterogeneous item relationship graph, where the relationships between items are determined by the types of behaviors.
Then HICG learns the meanings of distinct behaviors by utilizing Gate-GNN ~\cite{Li2016GatedGS} on the constructed heterogeneous graph. 
After that, we propose a heterogeneous information crossing module in HICG to model a user's current interests and his long-term preference. 
More specifically, this module models users' current interests and long-term preferences by utilizing an intra-behavior attention layer, which is used to capture the relationships between their previous behaviors and most recent actions.
In order to obtain a more precise representation of users' current interests, this module also applies an inter-behavior attention layer to characterize the relationships between different behavior types.
Finally, HICG combines current interests with long-term preferences to make recommendations.
%
%
In addition, inspired by the effectiveness of contrastive learning (CL) techniques in many areas ~\cite{Oord2018RepresentationLW,Yu2021SelfSupervisedMH}, we propose an enhanced version, named HICG-CL, which utilizes an extended CL module to improve the performance of HICG. 
This module, in particular, uses the item co-occurrence relationships across different sessions to build a union graph, and then applies a CL task on the graph to improve the learning of item representations.
HICG-CL combines HICG with the CL module in a unified architecture, and unifies task and CL task under a multi-task learning framework. 

The main contributions in this work are summarized as follows:
\begin{itemize}[leftmargin=*] \setlength{\itemsep}{-\itemsep}
    \item We propose a novel method HICG for session-based recommendation, which can utilize multiple types of user behaviors to capture users' current interests and their long-term preferences through heterogeneous graph modeling.
    \item We develop a heterogeneous information crossing module, which can effectively learn the relationships among the various behaviors with different user behavior types. 
    \item We also propose an enhanced version, named HICG-CL, which incorporates CL to enhance the learning of item representations.
    \item We conduct extensive experiments on three real-world datasets, and the state-of-the-art performance demonstrates the superiority of our proposed methods.
    %
\end{itemize}

The rest of this article is organized as follows.  
\autoref{ss:2} reviews the related work on existing session-based recommendation methods. 
\autoref{ss:3} presents our proposed HICG and HICG-CL in details. 
\autoref{ss:4} analyzes the results of extensive experiments we conducted on three real-world datasets. 
\autoref{ss:5} gives the conclusion of this article.

%% file: content/2_related_work/content.tex
\section{Related Work} \label{ss:2}

This section reviews the related work on SBRS, including SRBS with identical type behaviors modeling,
SRBS with heterogeneous behaviors modeling, and previous work which utilizes contrastive learning for SRBS.

%

\input{./content/2_related_work/1_sbrs.tex}
\input{./content/2_related_work/2_hb.tex}

\input{./content/2_related_work/3_cl.tex}

%% file: content/2_related_work/1_sbrs.tex
\subsection{Identical Type Behaviors Modeling for SRBS}

Conventional methods are mainly based on collaborative filtering (CF) techniques and Markov chain (MC) sequential models. 
CF based methods provide recommendation services based on similarity. 
For example, IKNN ~\cite{Sarwar2001ItembasedCF} first compares the items' co-occurrences in different sessions to determine the session similarity, and then recommends objects in the top-K similar sessions that the current user has not touched.
MC based methods assume that user's next behavior is only related to the behaviors of his current moment.
According to this assumption, researchers apply the first-order Markov chain to model user's transfer between various interacted items, based on which session-based recommendation can be made by simply computing the transition probabilities between items ~\cite{Shani2002AnMR,Zimdars2001UsingTD}.
FPMC ~\cite{Rendle2010FactorizingPM} combines CF and MC approaches in a hybrid recommendation framework,
and learns users' long-term preferences with short-term interests to predict the items in their next shopping baskets. 

Neural network based models ~\cite{Barkan2016ITEM2VECNI,Covington2016DeepNN} have been proposed to model the sequential behaviors in SBRS with great success. 
For example, GRU4Rec ~\cite{Hidasi2016SessionbasedRW} first employs recurrent network technology with Gated Recurrent Unit (GRU) to model item interaction sequences.
Then, GRU4Rec+ ~\cite{Tan2016ImprovedRN} further enhances the performance of GRU4Rec by introducing a data augmentation technique.
In addition, GRU4REC-DWell ~\cite{Bogina2017IncorporatingDT} improves GRU4Rec by applying the duration of each user behavior in sessions. 
However, the recurrent neural network based models fail to learn the associations between users' long-term behaviors and their current interests. 
To address this issue, researchers have investigated models based on attention mechanisms, which can capture the relationship between users' latest actions and their long-term behaviors. 
By applying an attention mechanism, NARM ~\cite{narm2017jing} computes the importance scores of each historical behavior. 
After that, the scores are used by NARM to calculate a weighted sum of behavior representations extracted by recurrent network, and the calculation result is used to represent the user's current intent.
STAMP ~\cite{Liu2018STAMPSA} emphasizes the importance of the last click by using a global attention network. 
A trilinear product decoder is also used to improve its expressive ability, which can describe the relationship between users' long-term preferences, short-term interests, and the candidate items.
Inspired by Transformer ~\cite{Vaswani2017AttentionIA}, SASRec ~\cite{Kang2018SelfAttentiveSR} learns the associations between different behaviors in sessions by applying a multi-head self-attention mechanism, which also achieves great improvement.
With consideration of collaborative information in sessions and dynamic presentation of items, CoSAN ~\cite{Luo2020CollaborativeSN} enhances the recommendation performance by investigating neighborhood sessions.

Recently, researchers have introduced Graph Neural Network (GNN) technology to model the complex relationships in graph-structured data, achieving excellent performance in a variety of tasks~\cite{Wu2019ACS}.
According to previous work on SRBS ~\cite{Wu2019SessionbasedRW,Xu2019GraphCS,Yu2020TAGNNTA}, the association between items can be used to construct directed graphs, and using GNN on those graphs can effectively generate session representations.
%
SR-GNN ~\cite{Wu2019SessionbasedRW} is a representation work among them. 
SR-GNN generates a directed graph of interactions for each session, where the edges represent the order of adjacent interactions within the given session.
After that, SR-GNN processes each session graph to obtain the behavioral representations with an attention mechanism.
%
%
GC-SAN ~\cite{Xu2019GraphCS} improves SR-GNN by introducing a self-attention mechanism, which can differentiate the contributions of distinct adjacent nodes in the graph.
By applying a global attention mechanism, TAGNN ~\cite{Yu2020TAGNNTA} calculates the contributions of users' previous interactions and uses them to enhance the representations of their current interests.
FGNN ~\cite{Qiu2019RethinkingTI} and SGAT ~\cite{Chen2020HandlingIL} utilise graph attention networks (GAT) to extract node associations with multi-head attention. 
Both of them achieve a good breakthrough by effectively aggregating neighbors' representations 
as the target representation in the session.

Although the above mentioned methods have made remarkable progress on SBRS, these approaches only model a single type of user interactions.
In contrast, by building a heterogeneous item relationship graph and learning the relationships among the user's heterogeneous actions, our proposed HICG in this paper utilizes various user behaviors to capture users' current interests and long-term preferences.

%% file: content/2_related_work/2_hb.tex
\subsection{Heterogeneous Behaviors Modeling for SRBS}

To improve recommendation performance, SRBS methods with heterogeneous behavior modeling employ a variety of user interaction types.
With the consideration of distinct relationship types between items, those methods can model users' preferences more effectively. 
Although some early recommendation studies ~\cite{Koren2008FactorizationMT,Li2018LearningFH,Ding2018ImprovingIR,Chen2020EfficientHC} have utilized the heterogeneous information of user interactions to improve performance of traditional recommendation tasks, only a few studies in SRBS have investigated the approaches to generating the representations from multi-behavior sessions.
Among them, MKM-SR ~\cite{Meng2020IncorporatingUM} enhances the representations of the current session by incorporating user micro-behaviors 
and it also involves KG embedding learning as an auxiliary task to promote the recommendation effects. 
MCPRN ~\cite{Wang2019ModelingMS} utilizes an RNN-based mixed channel purpose routing network to model the interacted items in the sequence, where different channels are used to learn the distinguish interests. 
After that, MCPRN generates the recommendation results by the integrated representation from all the channels.
MGNN-SPred ~\cite{Wang2020BeyondCM} applies auxiliary behaviors with target ones by constructing a multi-relational item graph, which greatly improves the effectiveness of predicting the user's subsequent target behavior. 
Although the aforementioned methods have improved recommendation performance in SRBS, they use heterogeneous behaviors without considering the underlying relationships between them.
%
%
Unlike these methods, our proposed HICG makes better use of heterogeneous behaviours and captures their inherent relationships.

%% file: content/2_related_work/3_cl.tex
\subsection{Contrastive Learning for SRBS}
Contrastive learning (CL) is one of the hot areas of recent scientific research, which belongs to the discriminative self-supervised learning method.
Compared with generative self-supervised learning (SSL) methods such as GAN ~\cite{Goodfellow2014GenerativeAN} and VAE ~\cite{Kingma2014AutoEncodingVB}, CL focuses more on learning the general characteristics of instances by distinguishing between similar 
and non-similar entities ~\cite{Sun2021SUGARSN}.
To date, CL has a lot of studies in computer vision and natural language processing. 
However, only a few studies investigated the efficacy of using this approach in SRBS. 
DHCN ~\cite{Xia2021SelfSupervisedHC} models the session-based data as a hypergraph and introduces CL as an auxiliary task to enhance the supergraph modeling ability, which maximizes the mutual information between the session representations learned by two different networks.
MHCN ~\cite{Yu2021SelfSupervisedMH} uses the aggregation of high-order user relations to enhance social recommendation. 
To alleviate the weakness of feature differences caused by the aggregation of different levels of information, 
MHCN utilizes CL to obtain the connectivity that maximizes the hierarchical mutual information.
COTREC ~\cite{Xia2021SelfSupervisedGC} generates subgraphs by randomly dropping edges in the session graph and treats homologous subgraphs as identical sessions in model learning.
To obtain a general representation of the given session, COTREC applies CL to differentiate these subgraphs.
Different from the above methods that use CL to augment session representations, HICG applies CL to enhance the learning of item representations by their co-occurrence relationships across different sessions.

\newpage

%% file: content/3_method/content.tex
\section{Proposed Approach} \label{ss:3}

In this section, we illustrate the architecture and components of our proposed HICG. 
%

\input{./content/3_method/1_pre.tex}

\input{./content/3_method/2_hicg.tex}

\input{./content/3_method/3_ssl.tex}

%% file: content/3_method/1_pre.tex

\subsection{Problem Formulation}
Before describing HICG and its structure, we first formulate the session-based recommendation problem.
A session, in SBRS, refers to behavior records generated during a continuous time period by a user.
Different service platforms choose slightly distinct duration for a single session based on the time a user connects to a service. 
For example, platforms such as Airbnb\footnote[1]{\ \href{https://www.airbnb.com/}{https://www.airbnb.com/}} and Taobao\footnote[2]{\ \href{https://www.taobao.com/}{https://www.taobao.com/}} regard the continuous behavior of mobile users within no more than 30 minutes as the same session ~\cite{Feng2019DeepSI,Mihajlo2018Airbnb}. 
Many search engines, on the other hand, regard interactive behaviors that occur in a single browser opening as records in an identical session.

Due to the fact that session-based recommendations are not aiming at specific application scenarios, we refer to objects such as commodities, content, and resources as ``item'' in this article. 
We use $I$ to stand for the set of items and $\Upsilon$ for the set of behavior types.
We set $\bm{s}_T$ as the total number of behaviors in this session, and use $\bm{s}\ =\ <b_1,\ b_2,\ ...,\ b_t,\ ...,\ b_{\bm{s}_T}>$ to represent the current session, where $b_t$ is the $t$-th user interaction,
consisting of item $i_{t} \in I$ and behavior type $\upsilon_t \in \Upsilon$.
Let $\tau$ be the target type of behavior that we want to predict, e.g., clicks. 
Besides, we also use suitable symbols to represent the model structure and significant intermediate calculation results in our algorithm. 
We summarize the main notations with their meanings in \autoref{tab:xxx}.
Then, we formulate the session-based recommendation problem as follows: 

\begin{problem}[Session-based Recommendation~\cite{Wu2019SessionbasedRW}]
Given user's historical behaviors in the current session $\bm{s}$, session-based recommender systems use the experience/models learned from the training data to predict $\widehat{y}_{\bm{s}, i}$, which means the likelihood that candidate item $i$ will be interacted with in user's next behavior. 
The final list of recommendations includes items with the top-K greatest $\widehat{y}_{\bm{s}, i}$.

\end{problem}

\begin{table}
    \caption{Important Notations.}
    \label{tab:xxx}
    \begin{tabular}{p{0.13\columnwidth} p{0.8\columnwidth}}
        \toprule
        Notation&Description\\
        \midrule
        $\bm{s}$& a single session record \\
        $b$& a user interactive behavior\\
        $\bm{s}_T$& the number of behavior in session $\bm{s}$ \\
        $\upsilon,\ \tau$& the types of user behavior where $\tau$ refers to the target type \\
        $i$& the identification of an specific item\\
        $\mathcal{I}, \Upsilon$ & the set of all candidate items and behavior types \\  
        $\bm{e_i}$ & the embedding represent vector of item $i$ \\  
        $\bm{p}_{\bm{s}}$ & the user preferences in session $\bm{s}$ \\ 
        $\bm{q}_{\bm{s}_{\upsilon}}$ & the user requirements status of $\upsilon$ in session $\bm{s}$ \\ 
        $\bm{h_{t}}$  & the represent vector of the $t$-th behavior in a specific session \\ 
        $\alpha_{t}, \alpha_{\upsilon}$  & the attention scores of the $t$-th behavior and the behaviors with $\upsilon$ type\\ 
        $\bm{W}, \bm{v}$   &  learnable matrix and vectors where we use $\bm{b}$ to represent bias vector\\ 
        $\widehat{y}_{\bm{s}, i}$ &  the possibility of item $i$ will be interacted by the user in the next behavior\\ 
        \bottomrule
    \end{tabular}
\end{table}

%% file: content/3_method/2_hicg.tex
\begin{figure}[htbp]
    \centering
    \includegraphics[width=1.00\textwidth,height=0.75\linewidth]{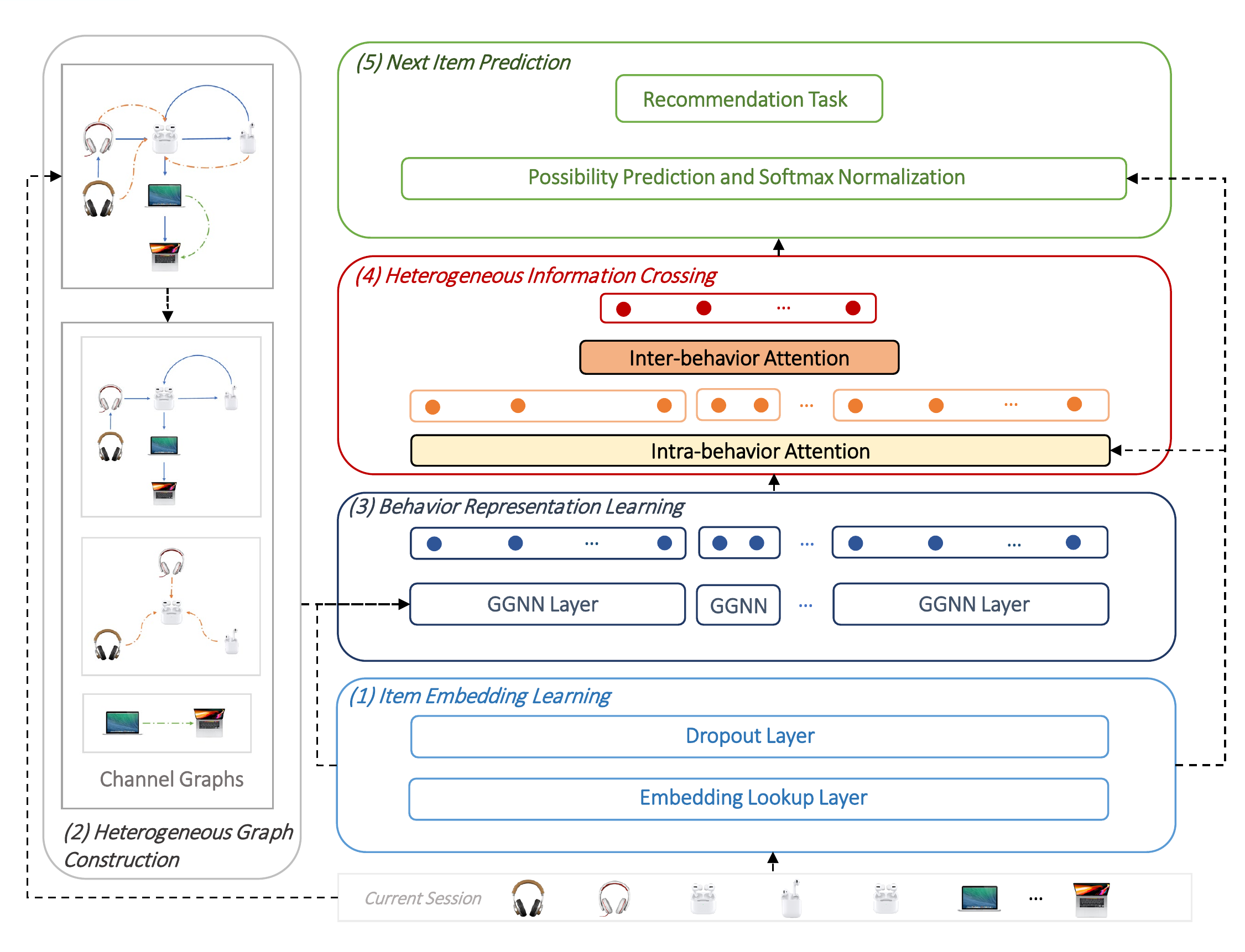}
    \caption{\label{fig:HICG}
    HICG model architecture.
    HICG contains five main components: 
    (1) Item Embedding Learning, 
    (2) Heterogeneous Graph Construction,
    (3) Behavior Representation Learning,
    (4) Heterogeneous Information Crossing,
    and (5) Next Item Prediction. 
    }
\end{figure}

\subsection{Heterogeneous Information Crossing on Graphs}

\input{./content/3_method/2_hicg/2_1_ov.tex}
\input{./content/3_method/2_hicg/2_2_iel.tex}
\input{./content/3_method/2_hicg/2_3_hgc.tex}
\input{./content/3_method/2_hicg/2_4_brl.tex}
\input{./content/3_method/2_hicg/2_5_hic.tex}
\input{./content/3_method/2_hicg/2_6_nip.tex}

%% file: content/3_method/2_hicg/2_1_ov.tex
\subsubsection{\textbf{Overview}}

To capture the relationships between different type of user behaviors, we propose a new graph-based method, named \textbf{H}eterogeneous \textbf{I}nformation \textbf{C}rossing on \textbf{G}raphs (HICG).
The overall structure of the proposed HICG is depicted in \autoref{fig:HICG}. 
There are five critical components in HICG, which are labeled by (1)-(5) in \autoref{fig:HICG}, and we give a brief description of them below:
\begin{itemize}[leftmargin=*] \setlength{\itemsep}{-\itemsep}
    \item \textbf{(1) Item Embedding Learning.} To effectively utilize the session context in HICG, we first use an embedding layer to project items to vector representations. 
    \item \textbf{(2) Heterogeneous Graph Construction.} Then, HICG builds a heterogeneous item relationship graph to model the different type of behaviors for a given session. 
    \item \textbf{(3) Behavior Representation Learning.} Next, Gate-GNN is used in HICG to learn the meanings of distinct behaviors on the constructed heterogeneous graph.
    \item \textbf{(4) Heterogeneous Information Crossing.} After that, HICG utilizes a heterogeneous information crossing module based on hierarchical attention mechanisms to model user's current interests and his long-term preference. 
    Specifically, in order to obtain user's current interests and characterize his long-term preferences, an intra-behavior attention layer is used to capture the correlations between user's historical behaviors and his most recent behaviors.
    By modeling the relationships between different behavior types, an inter-behavior attention layer is then used to improve the representation of user's current interests. 
    \item \textbf{(5) Next Item Prediction.} Finally, HICG uses the learned user and item representation to predict the next pontienial interacted item and make recommendation. 
\end{itemize}

We detail these modules as follows.

%% file: content/3_method/2_hicg/2_2_iel.tex
\subsubsection{\textbf{Item Embedding Learning}}
Firstly, we construct this module to transform the discrete inputs of sessions into dense vectors that HICG can handle.
Similar as the existing session-based recommendation models ~\cite{Hidasi2016SessionbasedRW,Xu2019GraphCS}, we apply an embedding layer to convert the items that a user has interacted with into vectors for each input session. 
Specifically, we use a one-hot vector $\bm{o}_i \in \mathbb{R}^{|\mathcal{I}|}$ to represent each item $i$ in the input data and transform it  to low dimensional dense vector $\bm{e}_i \in \mathbb{R}^{d}$  by a learnable matrix $\bm{W}_{emb} \in \mathbb{R}^{|\mathcal{I}| \times d} $: 

\begin{equation}
    \label{equ:3.1}
    \bm{e}_i \ = \ \bm{W}_{emb}^\top \bm{o}_i \ .
\end{equation}
\vspace*{0.5em}

We implement this module by a fully connected layer without bias terms and use the table look-up mechanism.
Besides, a dropout layer is used after obtaining those vectors when training HICG to avoid overfitting.

%% file: content/3_method/2_hicg/2_3_hgc.tex
\begin{figure}[htbp]
    \centering
    \includegraphics[width=0.8\linewidth]{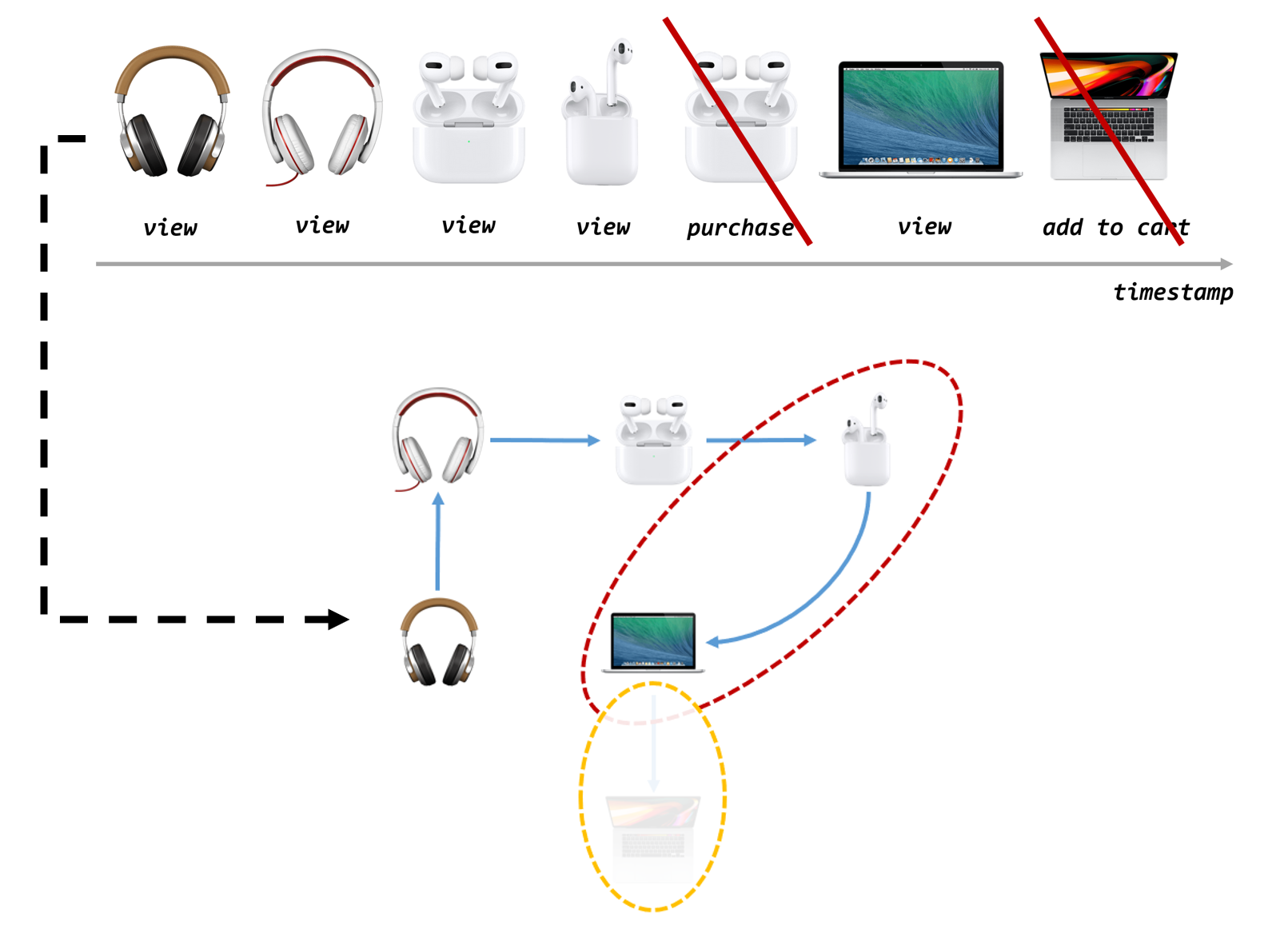}
    \caption{\label{fig:fig_3_1}An example to illustrate the potential shortcomings when SBRS faces a session like \autoref{fig:sample} and only considers a single type of behavior modeling.}
\end{figure}

\subsubsection{\textbf{Heterogeneous Graph Construction}} 
In practical application scenarios, user behaviors and their types are very rich. 
Different types of behaviors, such as browsing, adding-to-cart, purchasing, dwelling time, and ratings, represent users' different intents and can be utilized by SRBS to model users' preferences.

AS we discuss in \autoref{ss:2}, many existing research works, such as GRU4REC\cite{Hidasi2016SessionbasedRW}, NARM\cite{narm2017jing}, SR-GNN ~\cite{Wu2019SessionbasedRW}, use only one type of behavior data for modeling and ignore the variety of user behavior types.
However, ignoring to incorporate multiple types may result in a failure to capture a user's interest drifting behind this signal.
\autoref{fig:fig_3_1} shows the problem when only considering a single type of behavior modeling: the single-type modeling method is difficult to reasonably explain and characterize the user's interests changing behind the user interaction, from airpods to a macbook.
In addition, ignoring the last shopping cart behavior will also cause a bias in the estimation of the user's current interests.
Other works, such as MCPRN ~\cite{Wang2019ModelingMS} and MGNN-SPred ~\cite{Wang2020BeyondCM}, use heterogeneous behaviors to improve SRBS recommendation performance without taking into account the underlying relationships between them.

\label{sss:1}
\begin{figure}[htbp]
    \centering
    \includegraphics[height=0.57\linewidth]{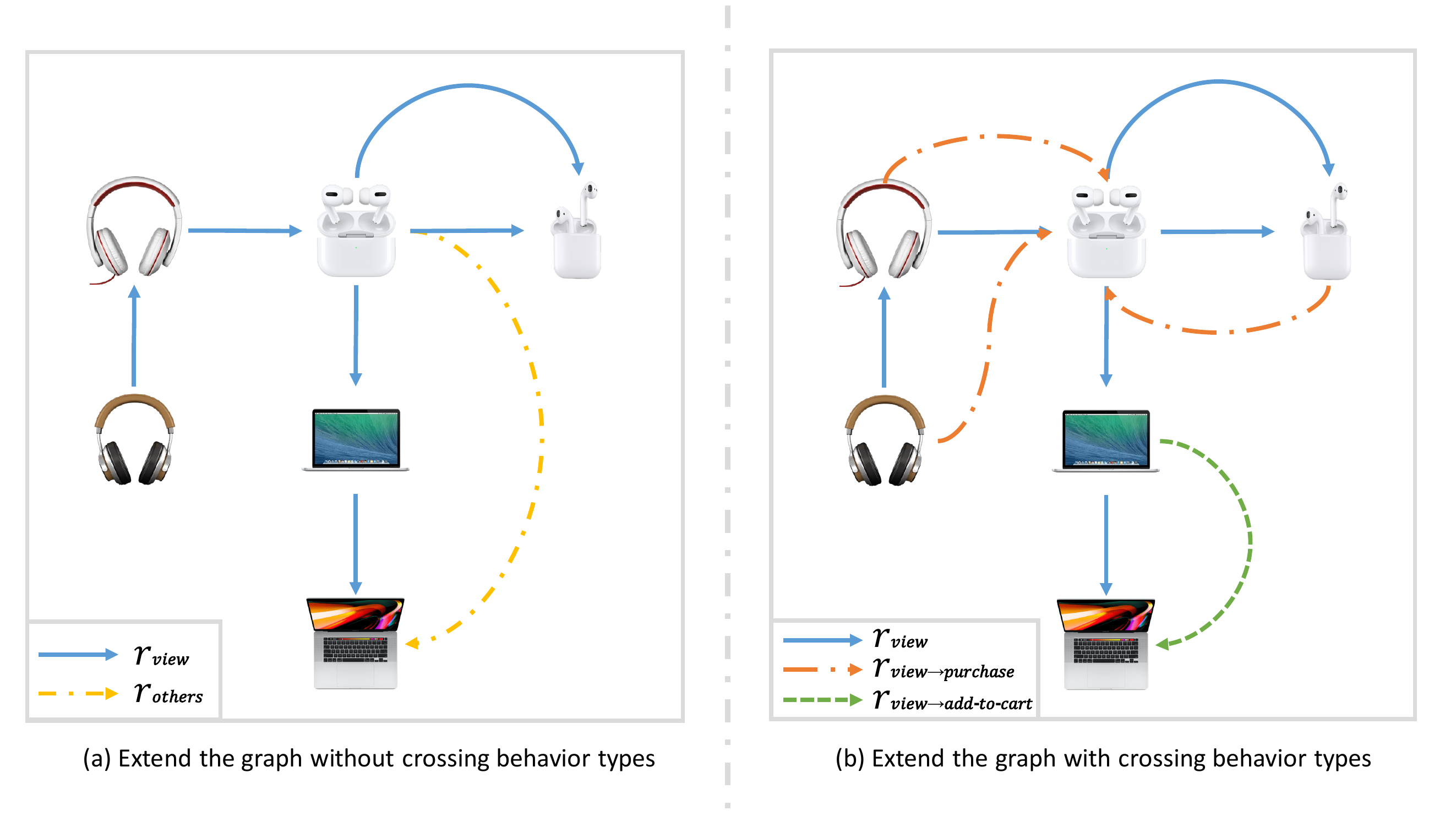}
    \caption{\label{fig:graph_construct} Heterogeneous graphs constructed by different approaches.
    %
    %
    (a) shows the heterogeneous graph constructed by``Extend the graph without crossing behavior type'' approach. 
    There are two types of edges in (a), i.e.,  
    the blue links are the $r_{view}$-type edges which connect two successively interacted items in a user's $view$-type behaviors, while the yellow links are the $r_{others}$-type edges that connect two successively interacted items in other type behaviors.
    (b) is constructed by``Extend the graph with crossing behavior type'', which contains three types of edges.
    The blue links have the same meaning as those in (a), the green and orange links represent the ``click''-to-``purchase'' and ``click''-to-``add-to-cart'' relationships between items, individually. 
    } 
\end{figure}

As there are multiple types of user behaviors in a given session, we construct a heterogeneous graph to utilize the relationships among them.
Specifically, we model each session $\bm{s}$ as a heterogeneous graph $\mathcal{G}_{\bm{s}} = (\mathcal{V}_{\bm{s}}, \mathcal{E}_{\bm{s}})$, where each node in $\mathcal{V}_{\bm{s}}$ represents an interacted item.
Each edge is a triplet $(a, b, r_{a, b}) \in \mathcal{E}_{\bm{s}}$, where the first two terms indicate that the user subsequently interacted with item $b$ after $a$, and the last term $r_{a, b}$ is an edge type that represents the relationships between these items.
The edge types depend on the types of user's behaviors.
For those target type ($\tau$-type) behaviors, such as views or clicks, we use $r_\tau$-type edges to connect items.
For other behaviors, we propose two approaches to construct edges below: \label{sss:3.2.3}

\begin{itemize}[leftmargin=*] \setlength{\itemsep}{-\itemsep}
    \item \textbf{Extend the graph without crossing behavior types.}
    Motivated by \cite{Wang2020BeyondCM}, we use an identical edge type $r_{others}$ to construct edges among the remained behaviors in this approach.
    The only distinction between (a, b, $r_{others}$) and (a, b, $r_\tau$) is that they are constructed by different user behavior types. 
    There is no intersection between behaviors of different types, which is the reason we call it ``extend the graph without crossing behavior types''. 
    \item \textbf{Extend the graph with crossing behavior types.} Some behaviors, such as purchases or long-term durations, can more accurately reflect the user's long-term preference, because they are related to previous behaviors while they indicate that the user's corresponding need has been met. 
    Consider the following situations from the perspective of users in session-based recommendation scenarios:
        \begin{itemize}
            \item[1.] In the e-commerce scenario, users browse and compare different products of the same type. 
            After that, the purchase or collection of such products occurs.
            \item[2.] In the video recommendation scenario, after briefly trying multiple videos of interest, the user is given a video selection to stay and watch.
            \item[3.] In the information retrieval scenario, after browsing and selecting multiple relevant information resources, users can either click in or end this search.
        \end{itemize}
    Motivated by this, we employ these behaviors to construct different types of edges between items. 
    Specifically, for a $\upsilon$-type behaviors $b_t$, we let $u$ denote $max(0,\ \mathop{argmax}_{t'}\mathbbm{1}\{\upsilon_{t'} \neq \tau \wedge t' < t \})$ and connect $i_t$ with all items $i \in \{i_{t'}\}_{u < t' < t}$ by edges with type of $r_{\tau \rightarrow \upsilon}$.
    As this approach utilizes different types of behaviors to construct edges, we name it ``Extend the graph with crossing behavior types.'' 
\end{itemize}

For the example session in \autoref{fig:sample}, we choose the $view$-type behavior as the target type $\tau$, and illustrate the difference between these heterogeneous graph construction approaches with two subfigures in \autoref{fig:graph_construct}. 

%% file: content/3_method/2_hicg/2_4_brl.tex
\subsubsection{\textbf{Behavior Representation Learning}}
To capture the connections between behaviors in the heterogeneous graph, we feed the graph into GNN to learn those behaviors' representations. 
As we focus on effectively modeling the behavior-level high-order relations among the user behaviors in this module, we apply different GNN networks to extract the representations of individual behaviors from the sub-graphs, which are generated from different channels according to the type of behaviors.
Similar to existing graph-based models ~\cite{Wu2019SessionbasedRW,Wang2020BeyondCM}, we use the degree matrices of the nodes in sub-graphs, 
as the matrices have a concise and clear expression form, and can be easily used for parallel computing.
On these sub-graphs, we apply the gate graph neural networks to extract the behavior representations. 
In this circumstance, the information of neighboring nodes is aggregated to the current node according to the products of the degree matrix and all involved node representations.
The update process of the gate graph neural network in HICG can be represented as follows:

\begin{equation}
    \label{equ:3.10}
    \bm{u}_{i_t} \ = \ \bm{A}_{i_t, :}[\bm{h}_1, ..., \bm{h}_N]^\top \bm{W}_G + \bm{b}  \ ,
\end{equation}

\begin{equation}
    \label{equ:3.11}
    \bm{z}_t \ = \ \sigma(\bm{W}_z \bm{u}_{i_t} + \bm{U}_z \bm{h}_t)  \ ,
\end{equation}

\begin{equation}
    \label{equ:3.12}
    \bm{r}_t \ = \ \sigma(\bm{W}_r \bm{u}_{i_t} + \bm{U}_r \bm{h}_t)  \ ,
\end{equation}

\begin{equation}
    \label{equ:3.13}
    \widetilde{\bm{h}_t} \ = \ \text{tanh} \Big(\bm{W}_o \bm{u}_{i_t} + \bm{U}_o (\bm{r}_t \odot \bm{h}_t) \Big)  \ ,
\end{equation}

\begin{equation}
    \label{equ:3.14}
    \bm{h}_t \ = \ (1 - \bm{z}_t) \odot \bm{h}_t + \bm{z}_t \odot \widetilde{\bm{h}_t}  \ ,
\end{equation}
where $\bm{A}$ is connection matrix which
is the concatenation of the out-degree and in-degree matrices of current graph,
and $A_{i_t, :}$ represents the row in $\bm{A}$ corresponding to the interacted item $i_t$ in the current session.
In addition, $\odot$ is an Hadamard product operator, $\sigma(\cdot)$ is the sigmoid function, 
$\bm{z}_t$ is an update gate, and $\bm{r}_t$ is a reset gate.
The initial input of $\bm{h}_t$ is the item embedding vector $\bm{e}_{i_t}$.

%% file: content/3_method/2_hicg/2_5_hic.tex
\subsubsection{\textbf{Heterogeneous Information Crossing}}
 
In order to effectively obtain the representations of the user's current interests $\bm{p}_{\bm{s}}$ and his long-term preference $\bm{q}_{\bm{s}_\upsilon}$ in the given session, we propose a heterogeneous information crossing approach with a hierarchical attention mechanism.
More specifically, we apply an intra-behavior attention layer to instruct the model care more about the relationships between historical behaviors and the latest interaction. 
By using this layer, we obtain the candidate representation of a user's current interests $\bm{p}_{\bm{s}}^{\ '}$ with his long-term preference. 
After that, HICG uses an inter-behavior attention layer to capture the relationships between different types of user interactions, and improves those representations. 
We detail the heterogeneous information crossing approach in the following:

\begin{itemize}[leftmargin=*] \setlength{\itemsep}{-\itemsep}
    \item \textbf{Intra-behavior Attention.} For each behavior type $\upsilon$, this module takes the corresponding users' behaviors representations as the input.
    These behavior representations, with the embedding vector of the latest interacted item, are applied to calculate attention scores by using a bilinear function.
    Finally, this module aggregates all of the representations to generate $\bm{p}_{\bm{s}}^{\ '}$ and $\bm{q}_{\bm{s}_\upsilon}$ by using a linear function.
    The process can be formulated as follows:
    \begin{equation}
        \label{equ:3.4}
        \alpha_t  \ = \ \frac{\text{exp} (\bm{e}_{\bm{s_T}}^\top \bm{W}_{\upsilon} \bm{h}_t)}
        {\sum_{t \in \mathcal{I}_{\bm{s}_{\upsilon}}}
         \text{exp} (\bm{e}_{\bm{s_T}}^\top \bm{W}_{\upsilon} \bm{h}_t)}  \ ,
    \end{equation}
    
    \begin{equation}
        \label{equ:3.5}
        \bm{p}_{\bm{s}}^{\ '} \ = \ \sum_{t \in \mathcal{I}_{\bm{s}_{\tau}}} \alpha_t \bm{h_{t}} \ , \quad
        \bm{q}_{\bm{s}_\upsilon} \ = \ \sum_{t \in \mathcal{I}_{\bm{s}_{\upsilon}}} \alpha_t \bm{h_{t}}   \ ,
    \end{equation}

    where $\bm{p}_{\bm{s}}^{\ '}$ is the candidate user interests in the current session, which is the weighted sum of his target behavior representation vectors,  and $\bm{q}_{\bm{s}_\upsilon}$ represents the user's long-term preference according to his behaviors with the type of $\upsilon$.

    \item \textbf{Inter-behavior Attention.} After obtaining the representations vectors for each type of user's historical behaviors, we use an inter-behavior attention layer to capture the relationships among them. 
    Specifically, HICG involves the representation vectors of different behavior types $\upsilon$ to calculate their attention scores with the candidate user preference vector, 
    and aggregates them with their scores to obtain $\bm{p}_{\bm{s}}$. 
    The corresponding formulae are as follow:
    \begin{equation}
        \label{equ:3.7}
        \alpha_\upsilon \ = \ \frac{\text{exp} \Big(\bm{v}^\top (\bm{W}_{\tau} \bm{p}_{\bm{s}}^{\ '}  + \bm{W}_{\upsilon} \bm{q}_{\bm{s}_\upsilon} + \bm{b}) \Big)}
            {\sum_{\upsilon} \text{exp} \Big(\bm{v}^\top (\bm{W}_{\tau} \bm{p}_{\bm{s}}^{\ '}  + \bm{W}_{\upsilon} \bm{q}_{\bm{s}_\upsilon} + \bm{b}) \Big)}  \ ,
    \end{equation}

    \begin{equation}
        \label{equ:3.8}
        \bm{p}_{\bm{s}} \ = \ \bm{p}_{\bm{s}}^{\ '} + \sum_{\upsilon} \alpha_\upsilon \bm{q}_{\bm{s}_\upsilon} \ ,
    \end{equation}
    where we calculate the attention score $\alpha_\upsilon$ by the soft-attention approach,
    and $\bm{p}_{\bm{s}}$ is the final representation of the user's current interests.
\end{itemize}

%% file: content/3_method/2_hicg/2_6_nip.tex
\subsubsection{\textbf{Next Item Prediction}}
After obtaining both of the item embeddings and user representations, we make predictions directly in this component, as HICG is designed for session-based recommendation tasks.
We calculate the possibility $\widehat{y}_{\bm{s}, i}$ that item $i$ be interacted with in a user's next behavior.
To be more specific, for each candidate item $i$, we first compute a candidate score.
Then we apply a softmax function to get the possibility predictions by normalizing the candidate score.
These processes can be defined as:
\begin{equation}
    \label{equ:3.2}
    \widehat{y}_{\bm{s}, i}^{\ '}  \ = \ \bm{e}_i^\top \bm{p}_{\bm{s}} - \bm{e}_i^\top \bm{W}\sum_{\upsilon}\bm{q}_{\bm{s}_\upsilon}  \ ,
\end{equation}

\begin{equation}
    \label{equ:3.3}
    \widehat{y}_{\bm{s}, i}  \ = \ \frac{\text{exp} ( \widehat{y}_{\bm{s}, i}^{\ '} )}{\sum_{i \in \mathcal{I_C}} \text{exp} ( \widehat{y}_{\bm{s}, i}^{\ '})}  \ ,
\end{equation}
where $\mathcal{I_C}$ is the set of candidate items. 

We explain the design motivation of \autoref{equ:3.2} as follows. 
First, to calculate the correlations between a candidate item $i$ and the user's current interests, we multiply its embedding $\bm{e}_i$ by the user's interests representation $\bm{p}_{\bm{s}}$, i.e., $\bm{e}_i^\top \bm{p}_{\bm{s}}$.
Second, as we mentioned in \autoref{sss:3.2.3}, the behaviors of type $\upsilon$, e.g., purchases, reflect the user’s long-term preference, since they indicate the corresponding needs have been satisfied. 
Third, we subtract $\bm{e}_i^\top \bm{W}\sum_{\upsilon}\bm{q}_{\bm{s}_\upsilon}$ from $\bm{e}_i^\top \bm{p}_{\bm{s}}$ to make users pay less attention to the items which they are already satisfied with.
%

%
Finally, items with the highest predicted possibilities will be recommended to the user.
As the goal of this module is to predict the correct item interacted with by the user next to the recommendation, we let HICG minimise the cross-entropy loss (CELoss) $\mathcal{L}_{HICG}$ between ground-truth and predictions, which can be defined as follows:

\begin{equation}
    \label{equ:3.15}
    \mathcal{L}_{HICG} \ = \ - \sum_{i \in \mathcal{I}} 
        y_{\bm{s}, i} \ \text{log}\widehat{y}_{\bm{s}, i} + 
        (1 - y_{\bm{s}, i}) \ \text{log}( 1 - \widehat{y}_{\bm{s}, i} )  \ .
\end{equation}

%% file: content/3_method/3_ssl.tex
\subsection{Enhancing HICG with Contrastive learning}
Although only the behaviors in the current session can be utilized to predict the user's next behavior in SBRS, we can still model the items' potential relationships as they may co-occur in different sessions.
HICG properly utilizes heterogeneous information through Heterogeneous Graph Construction and Heterogeneous Information Crossing modules. 
But except for the sharing embeddings of items in different sessions, it is still hard to utilize the item co-occurrence relationships across sessions. 
Contrastive learning, as a way of distinguishing between homogeneous and non-homogeneous entities by maximizing mutual information, enhances the embedding representations of similar entities that are relatively close in vector space, while the spatial distance between dissimilar entities is greater.
The co-occurrence relationships of items in different sessions can be used to measure the similarity among items, which motivates us to introduce a contrastive learning task to enhance HICG in this section.

The key to applying contrastive learning technology is how to combine application scenarios to construct meaningful positive and negative sample pairs.
In the existing work of session-based recommendation algorithms, as we discuss in \autoref{ss:2}, contrastive learning usually constructs positive sample pairs by means of different representations of the same session in multiple modalities, or random dropout of connecting edges in the item-user bipartite graph, etc. 
Although, from the actual effect, these construction methods have indeed brought a certain improvement to the prediction effect of the session recommendation algorithm, there are still some shortcomings: For starters, these works are based on contrastive learning of session-granular representations, which results in the learned representation vector not being able to transfer between sessions.
Second, these research efforts still lack the capture and application of the cross-conversational linkages between diverse items.

\begin{figure}[htbp]
    \centering
    \includegraphics[width=1\linewidth,height=0.6\linewidth]{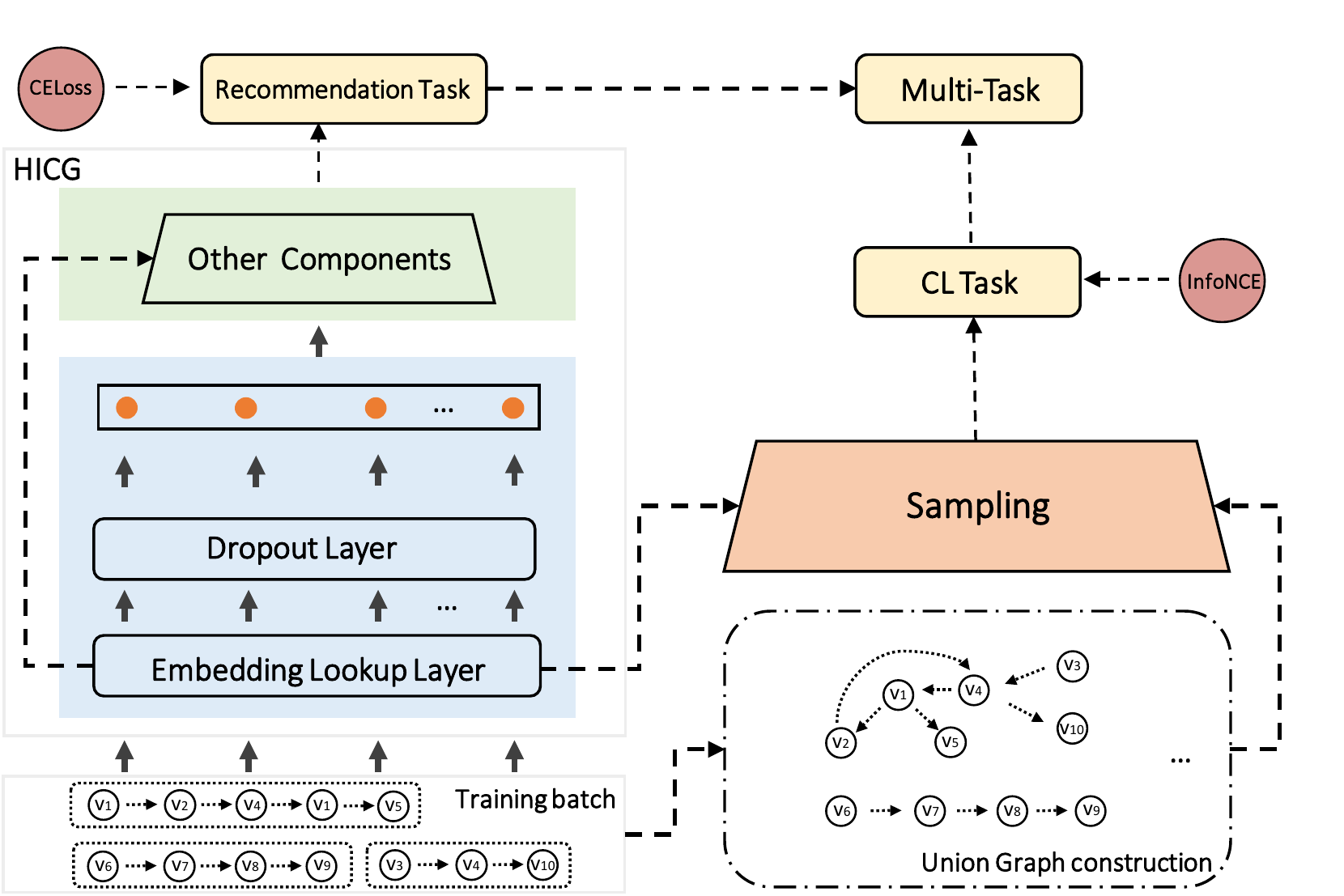}
    \caption{The framework of HICG-CL.}
    \label{fig:HICG-SSL}
\end{figure}

Following the recent SRBS works ~\cite{Xia2021SelfSupervisedHC,Yu2021SelfSupervisedMH}, correlated contrastive learning (CL) tasks can enhance the representation learning and lift the models' performance. 
With the motivation of using items' co-occurrence relationships across different sessions, we adopt this technique to enhance the learning of items' representations.
For simplicity, we call the CL enhanced version of HICG as HICG-CL.
HICG-CL mainly has the following two steps:
\begin{itemize}[leftmargin=*] \setlength{\itemsep}{-\itemsep}
    \item \textbf{Union graph construction}. Similar to the heterogeneous graph construction module in HICG, HICG-CL first constructs graphs for all sessions in the current training batch. 
    Then HICG-CL merges those graphs into one union graph, and the sessions that share the same items will be merged into a connected component of it.
    \item \textbf{CL pairs generation}. After constructing the union graph, HICG-CL applies a sampling module to construct CL pairs from the graph.
    Specifically, the positive pairs are generated by two items in the same connected component in the union graph.
    For each connected component $\mathcal{P}$, the sampling module randomly samples a set of negative items $\mathcal{X}_{\mathcal{P}}^-$ from other connected component with a ratio of $\beta$, and the negative pairs are generated as $(x, x^{-})$, where $x \in \mathcal{P}$ and $x^{-} \in \mathcal{X}_{\mathcal{P}}^-$.
\end{itemize}

We apply InfoNCE~\cite{Oord2018RepresentationLW} as an optimize target $\mathcal{L}_{CL}$ for the contrastive learning task, which can be defined as follow:
\begin{equation}
    \label{equ:4.3}
    \begin{aligned}
         &\ \mathcal{L}_{CL}(\mathcal{S})  \ = \ \sum_{\mathcal{P} \in \mathbb{P}_{\mathcal{S}}} \mathcal{L}_{CL}(\mathcal{P}, \mathcal{X}_{\mathcal{P}}^-;\ \mathcal{S})\\
        = &\sum_{\mathcal{P} \in \mathbb{P}_{\mathcal{S}}} - \frac{1}{|\mathcal{P}|} \sum_{(x, y) \sim \mathcal{P}} \text{log} \frac{\text{exp}(\text{sim}(\bm{e}_x, \bm{e}_{y}) / \tau)}
    {\sum_{x' \in \{y\} \cup \mathcal{X}_{\mathcal{P}}^-} \text{exp}(\text{sim}(\bm{e}_x, \bm{e}_{x'}) / \tau) } \ ,
    \end{aligned}
\end{equation}
where $\mathcal{S}$ is the set of sessions in current batch and $\mathbb{P}_{\mathcal{S}}$ it the set of connected components in the union graph. 
$(x, y)$ is a positive item pair where both $x$ and $y$ are sampled from the same connected component $\mathcal{P}$, and $\mathcal{X}_{\mathcal{P}}^-$ is the sampled negative item set for $\mathcal{P}$.
$\tau$ is a temperature parameter and $\text{sim}: \mathbb{R}^d \times \mathbb{R}^d \mapsto \mathbb{R}$ is the similarity function and we use cosine in this paper.

The framework of HICG-CL is shown as \autoref{fig:HICG-SSL}, in which we joint the contrastive learning task with the next item prediction task under a multi-task learning (MTL) framework. 
When the model is in the training process, the sessions involved in a training batch compose $\mathcal{S}$, which generates a corresponding connected component $\mathcal{P}$ and a sampled negative item set $\mathcal{X}_{\mathcal{P}}^-$ for the contrastive learning of item representation vectors.
The whole optimized target of HICG-CL is shown below. 
\begin{equation}
    \label{equ:4.5}
    \mathcal{L}_{HICG-CL} \ = \ \mathcal{L}_{HICG} + \lambda_{CL} \  \mathcal{L}_{CL}. 
\end{equation}

%% file: content/4_experiments/content.tex
\section{Experiments} \label{ss:4}

In this section, we conduct extensive experiments on three real-world recommendation datasets to evaluate the performance of HICG and analyze it. 
The purpose of this section is to answer the following research questions.

\begin{itemize}[leftmargin=*] \setlength{\itemsep}{-\itemsep}
    \item RQ1: Whether the proposed approach HICG outperforms the state-of-the-art session-based recommendation methods?
    \item RQ2: How does the heterogeneous information crossing affect the performance of HICG? 
    \item RQ3: How about the influence of different construct approaches on the recommendation? 
    \item RQ4: How about the performance of HICG facing different lengths of behavior sequences?
    \item RQ5: How the contrastive learning task affects the performance of HICG-CL?
    \item RQ6: How about the trainability of HICG than other state-of-the-art methods? 
    \item RQ7: How about the practicality of the HICG algorithm?
\end{itemize}

\input{./content/4_experiments/su.tex}
\subsection{Results Analysis}

\input{./content/4_experiments/ra_1.tex}

\input{./content/4_experiments/ra_2.tex}

\input{./content/4_experiments/ra_3.tex}

\input{./content/4_experiments/ra_4.tex}

\input{./content/4_experiments/ra_6.tex}
\input{./content/4_experiments/ra_5.tex}

\input{./content/4_experiments/ra_7.tex}

%% file: content/4_experiments/su.tex
\subsection{Experimental Setup}

\subsubsection{Datasets}
We evaluate the proposed model on three real-world recommendation system datasets,
i.e., Yoochoose\footnote[1]{\ \href{http://2015.recsyschallenge.com/challege.html}{http://2015.recsyschallenge.com/challege.html}}, Diginetica\footnote[2]{\ \href{http://cikm2016.cs.iupui.edu/cikm-cup}{http://cikm2016.cs.iupui.edu/cikm-cup}} and Retailrocket\footnote[3]{\ \href{https://www.kaggle.com/retailrocket/ecommerce-dataset}{https://www.kaggle.com/retailrocket/ecommerce-dataset}}.
All of these datasets are published by contests which contain many attributes of items and heterogeneous user-item interactions, e.g.,click, view, add-to-cart, and purchase.
%
%
In these interactions, Views (or clicks in Diginetica) are considered $\tau$-type behaviors, while other interactions are taken as users' other types of behaviors.
We follow the previous work ~\cite{Hidasi2016SessionbasedRW,Xu2019GraphCS} to preprocess these datasets for a fair comparison.
Specifically, we take the behaviors with the same session identity as a session directly in Yoochoose and Diginetica, and take the continuous user behaviors in 30 minutes as a session in Retailrocket.
%
For Yoochoose, we choose the sessions from the last day as test data. While for other datasets, we use the sessions from the most recent week as test data. 
Then we keep the sessions that contain at least two user interactions and filter out all items that have been browsed fewer than five times. 
After that, we follow ~\cite{narm2017jing,Wu2019SessionbasedRW} and construct training samples by using a sequence splitting method.
For a session $\bm{s} = <b_1, b_2, ..., b_{\bm{s}_T}>$, we generate $s_{T}-1$ samples by the behavior subsequences $<b_1>$, $<b_1,b_2>$, ...,$<b_1,b_2,...,b_{s_{T}-1}>$ and the next behavior as their labels $b_2,b_3,...,b_{\bm{s}_T}$.
At last, the Yoochoose dataset is quite large, and ~\cite{Tan2016ImprovedRN} shows that the performance obtained is more reliant on the recent sessions data. 
Thus, we only use the recent fractions of 1/64 train samples in Yoochoose. 
The statistics of these evaluation datasets after preprocessing are summarized in \autoref{tab:datasets_stat}.

\begin{table}[ht]
    \centering
    \renewcommand\arraystretch{1.3}
    \begin{tabular}{lrrr}
        \toprule
        \textbf{Dataset}   & \textbf{Yoochoose} & \textbf{Diginetica} & \textbf{Retailrocket} \\ 
        \midrule
        \# of train sessions & 369,859  & 719,470  & 749,947  \\
        \# of test sessions & 55,898  & 43,097  &  28,445 \\
        \# of items & 16,766 & 60,858 & 48,989 \\
        \# of views &  557,248  & 982,961 & 1,085,217 \\
        \# of conversions & 19,787 & 22,026  & 58,997  \\
        average length of views & 6.16 & 5.12  & 6.82 \\
        average length of conversions &  0.31 &  0.19  & 0.51 \\
        \bottomrule
    \end{tabular}
    \caption{\label{tab:datasets_stat} Statistics of the evaluation datasets.}
\end{table}

\subsubsection{Baseline Methods}
We compare the proposed model with several representatives or state-of-the-art methods for the session-based recommendation, 
which can be divided into two categories according to whether they utilize heterogeneous user behaviors or not. 
The first category of baselines contains methods dealing with single-type behaviors, which includes the following:

\begin{itemize}[leftmargin=*] \setlength{\itemsep}{-\itemsep}
    \item \textbf{S-POP}.\; It always recommends the most popular items in the current session.
    \item \textbf{IKNN}~\cite{Sarwar2001ItembasedCF}. This method recommends the items in the most parallel sessions where 
    the similarity is defined as the co-occurrence number of interacted items.
    \item \textbf{BPR-MF}~\cite{Rendle2009BPRBP}.  
    It is a representative method that uses a pairwise ranking loss to learn the parameters in the matrix factorization model.
    \item \textbf{GRU4Rec}~\cite{Hidasi2016SessionbasedRW}. It utilizes GRU layers to construct a sequential model for the session-based recommendation.
    \item \textbf{NARM}~\cite{narm2017jing}. A RNN-based model employs an attention mechanism to capture the relationship between the most recent behavior and the former.
    \item \textbf{STAMP}~\cite{Liu2018STAMPSA}. It uses self-attention layers to capture both the user's long-term preference and current interest to predict the next interaction.
    \item \textbf{SR-GNN}~\cite{Wu2019SessionbasedRW}. This method applies a gated graph neural network 
    to model the transitions of user interactive items in the given session.
\end{itemize}

The second category of baselines employ heterogeneous behaviors 
in the sessions are as follows:
\begin{itemize}[leftmargin=*] \setlength{\itemsep}{-\itemsep}
    \item \textbf{MCPRN}~\cite{Wang2019ModelingMS}. It models a session with a mixture-channel recurrent network and integrates all the channels to rank the candidate items.
    \item \textbf{MGNN-SPred}~\cite{Wang2020BeyondCM}. This method builds multiple graphs based on different types of behaviors and uses the candidate items concatenate embedding in different graphs to predict the next one interacted.
\end{itemize}

\subsubsection{Evaluation Metrics}
Following~\cite{narm2017jing,Wu2019SessionbasedRW}, we adopt two popular metrics
to evaluate the session-based recommendation performance of all models as follows,
\begin{itemize}[leftmargin=*] \setlength{\itemsep}{-\itemsep}
    \item \textbf{HR@K} (Hit Ratio) measures the ratio of correctly recommended items among top-K of the ranking list
    where $N$ is the number of sessions used for evaluation, and $n_{hit}$ represents the number of hits of the algorithm.
    \begin{equation}
        \label{equ:3.3.1}
        HR@K \ = \ \frac{n_{hit}}{N} \ .
    \end{equation}
    \item \textbf{MRR@K} (Mean Reciprocal Rank), evaluates the quality of the rank of ground-truth in the top-K candidate items list.
    The set $\mathcal{T}_i$ represents the candidate items set, and $rank(t,\ l_i)$ represents the rank of the target object in the list $l_i$.
    \begin{equation}
        \label{equ:3.3.2}
        MRR@K \ = \ \frac{1}{N} \sum_{i=1}^{N} \sum_{t \in \mathcal{T}_i}\ \frac{1}{rank(t,\ l_i)}  \ .
    \end{equation}
\end{itemize}
This study considers the top-K (K = 20, 5) results with the highest prediction scores in the whole set of all items as the candidates.

\subsubsection{Hyper-parameter settings}
In the comparative experiments for the session-based recommendation algorithms, including the proposed HICG, there are some hyperparameters that need to be set.
We implement all of the baselines and tune all parameters via 
grid search on each dataset to achieve optimal performances for a fair comparison.
Specifically, in the following experiments reported, we set the dimension of item embedding $d$ as 100. 
The dropout ratio is set to 0.2 in the dropout layer. 
The number of channels in MCPRN is 3.
%
The negative sampling ratio $\beta$ in our proposed model is set to 0.1 for Diginetica, and 0.2 for other datasets. 
We set the magnitude of the contrastive learning task $\lambda_{CL}$ to 0.1 for all datasets. 
We also design an experiment to analysis the influence of $\lambda_{CL}$ and $\beta$. 
Adam is used as the model optimizer with the mini-batch size is set to 100.
The initial learning rate is set to $3\times 10^{-4}$, and the $L_2$ regularization is $10^{-5}$. 

%% file: content/4_experiments/ra_1.tex
\subsubsection{Overall Performance (RQ1)}
To answer RQ1, we compare our proposed model HICG with other state-of-the-art session-based recommendation methods to demonstrate its performance.
The second approach in \autoref{sss:3.2.3} is chosen to construct the heterogeneous graph in this experiment.
The overall performance in terms of HR@20 and MRR@20 on three datasets are shown in shows in \autoref{tab:Performance_comparison_1}. 
From the results, we can have the following observations:

\begin{table}[htbp]
    \centering
    \renewcommand\arraystretch{1.3}
    \begin{tabular}[3\columnwidth]{cccccccc}
        \toprule
        \vspace*{-0.3em}
        & & \multicolumn{2}{c}{\textbf{Yoochoose}} 
        & \multicolumn{2}{c}{\textbf{Diginetica}} 
        & \multicolumn{2}{c}{\textbf{Retailrocket}} \\
        \multicolumn{2}{c}{\textbf{Algorithm}}
        \vspace*{0.3em}
        &\multicolumn{2}{c}{\rule{0.23\columnwidth}{0.1pt}} 
        &\multicolumn{2}{c}{\rule{0.23\columnwidth}{0.1pt}} 
        &\multicolumn{2}{c}{\rule{0.23\columnwidth}{0.1pt}} \\ 
        & &\textbf{HR@20} &\textbf{MRR@20} &\textbf{HR@20} &\textbf{MRR@20} &\textbf{HR@20} &\textbf{MRR@20} \vspace*{0.1em}\\
        \midrule 
        & S-POP & 30.44 & 18.35 & 21.06 & 13.68 & 38.03 & 24.81 \\
        & IKNN & 51.60 & 21.81 & 35.75 & 11.57 & 40.68  & 31.05 \\
        & BPR-MF & 31.31 & 12.08 & \hspace*{0.3em} 5.24 & \hspace*{0.3em} 1.98 & 31.75 & 19.40 \\
        \textbf{Single} 
        & GRU4Rec & 60.64 & 22.89 & 29.45 & \hspace*{0.3em} 8.33 & 44.01 & 23.67 \\
        & NARM & 68.32 & 28.63 & 49.70 & 16.17 & 50.22 & 25.38 \\
        & STAMP & 68.74 & 29.67 & 45.64 & 14.32 & \underline{50.96} & 25.17 \\
        & SR-GNN & \underline{70.57} & \underline{30.94} & 50.73 & 17.59 & 50.32 & 26.57 \\
        \midrule
        \multirow{2}*{\textbf{Heter.}}  
        & MCPRN  & 68.33 & 29.85  & 46.35 & 16.04  & 49.28 & 24.56 \\
        ~ & MGNN-SPred  & 69.36 & 29.48  & \underline{50.91} & \underline{18.27}  & 50.94 & \underline{28.45} \\
        \midrule
        \multirow{2}*{\textbf{Ours}} & \textbf{HICG} & 73.06 & \textbf{31.59}  & 51.03 & 19.77  & 52.87 & 29.15 \\
        ~ & \textbf{HICG-CL} & \textbf{73.19} & 31.27  & \textbf{51.21} & \textbf{20.51}  & \textbf{52.92} & \textbf{29.30} \\
        \bottomrule
    \end{tabular}
  
   \caption{\label{tab:Performance_comparison_1} 
    Comparison results on three datasets. 
    The best performance is shown in bold font, while the runner-up is shown in underline.
    }
\end{table}

\begin{itemize}[leftmargin=*] \setlength{\itemsep}{-\itemsep}
    \item \textbf{Similar items may repetitively co-occur in the same session.} 
    Although the first three conventional baselines S-POP, IKNN and BPR-MF perform worse than other neural-network-based methods, they still obtain significant results. 
    The S-POP method recommends the most popular items in the current session and achieves 38.03\% in Retailrocket, and the IKNN works better than GRU4Rec in Diginetica. 
    Both of those results show that users browse similar (even identical) items repetitively in their sessions, and they also indicate that the approach of using graphs to construct relationships between items makes sense.
    \item \textbf{Methods based on graph neural networks have more advantages than others.} 
    From the observations of the metrics of each algorithm on the experimental datasets, the SRBS algorithms based on deep learning generally have more advantages than the traditional method except for a few items.    
    Furthermore, algorithms based on the graph neural network (SR-GNN, MGNN-SPred, HICG) are better than those only applying the recurrent networks (GUR4Rec, NARM) or attention mechanism (STAMP).
    Besides, compared with GRU4REC, NARM has a more remarkable improvement by using the attention mechanism, while STAMP using only the attention mechanism has shown superior recommendation effects on multiple metrics.
    These results also show the rationality and validity of introducing the attention mechanism in SRBS. 
    Compared with NARM, which applies sequence-structure modeling, SR-GNN uses a graph network to replace the original sequence structure and achieves better results.
    \item \textbf{Methods using heterogeneous behaviors data generally work well than those only employ a single type.}
    The heterogeneous methods, MCPRN and MGNN-SPred, achieve better performance than single-type behavior-based methods, except SR-GNN, in almost all metrics. 
    Therefore, it is necessary for us to utilize heterogeneous information.
    \item \textbf{The proposed method HICG outperforms others baselines consistently on all datasets.} 
    The comparison results demonstrate the effectiveness and superiority of HICG and verify that the approach to utilizing heterogeneous behaviors is essential.
    Besides, HICG-CL also achieves an improvement by applying an additional contrastive learning task to enhance the learning of item representations.
\end{itemize}

\begin{figure}[ht]
    \centering
    \includegraphics[width=0.47\linewidth]{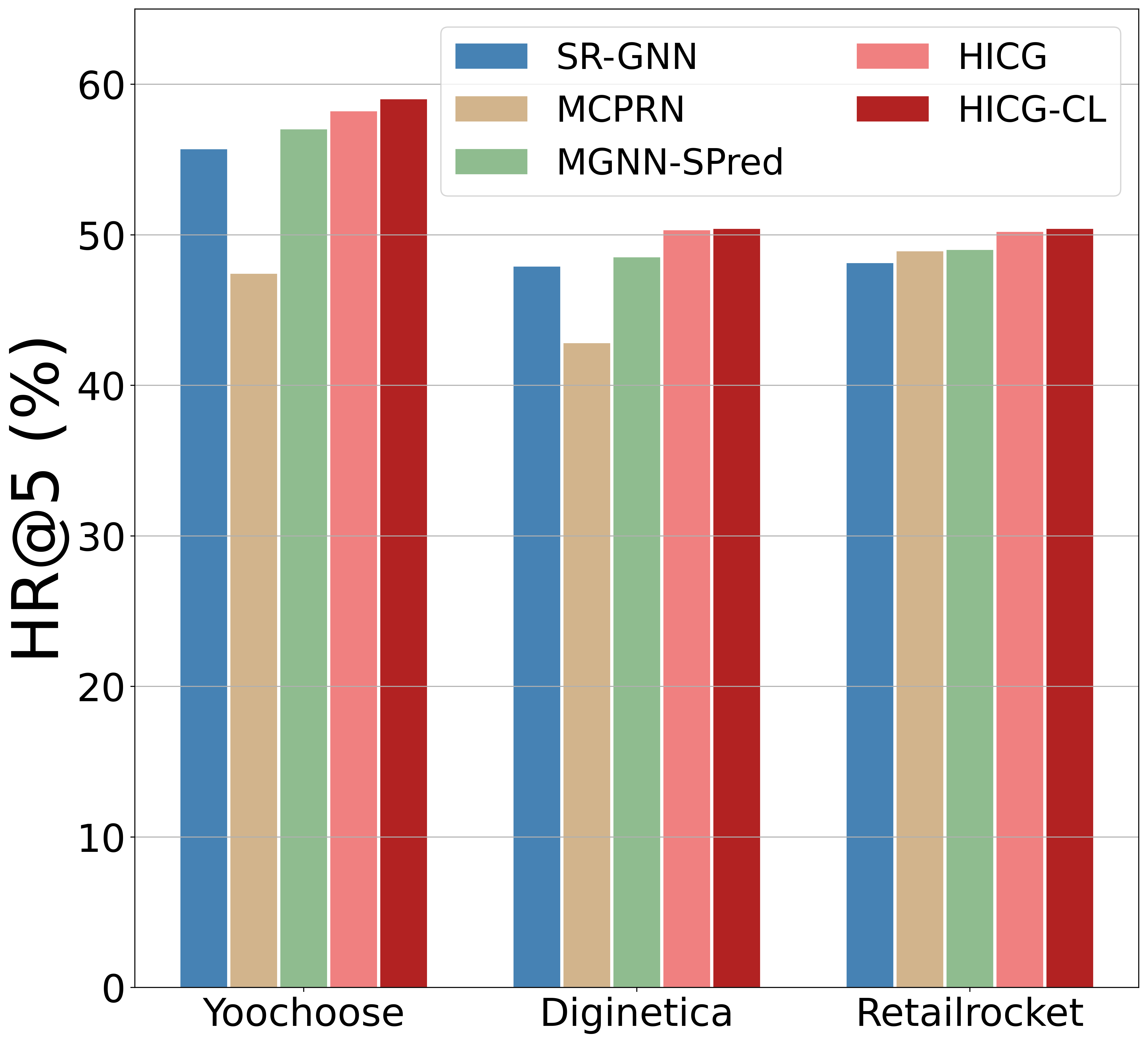}
    \qquad
    \includegraphics[width=0.47\linewidth]{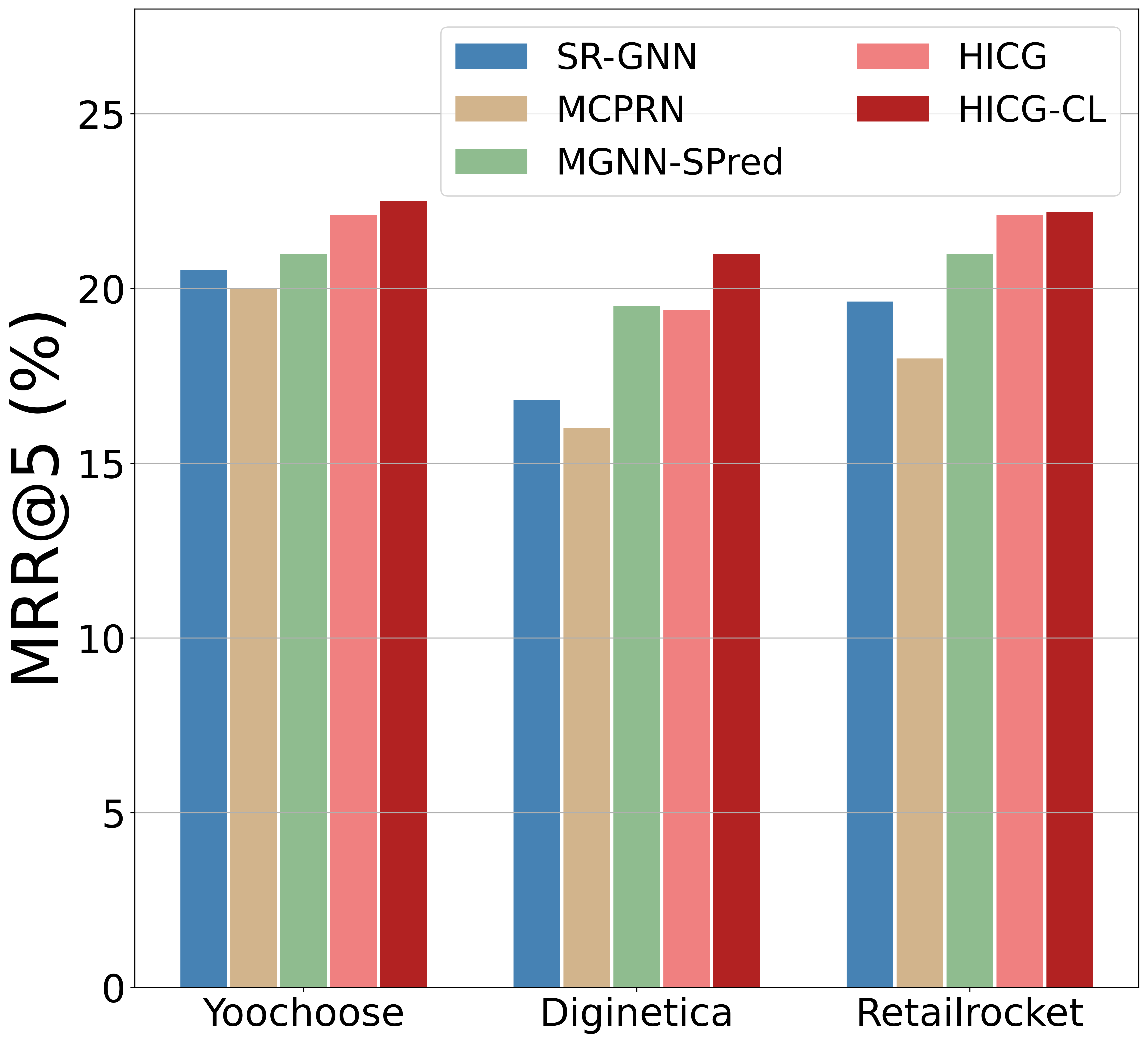}
    \caption{Performance of HICG and HICG-CL with K = 5.}
    \label{fig:fig5_1}
\end{figure}

We also verify the performance of HICG with higher quality requirements by using K = 5, and the results are shown in \autoref{fig:fig5_1}.
It can be seen from the experimental results that HICG still maintains a superior effect in more refined recommendations.

%% file: content/4_experiments/ra_2.tex
\subsubsection{Ablation Study (RQ2)}
We conduct an ablation experiment to check out the improvement lift
by the heterogeneous information crossing. 
The experimental result shown in \autoref{tab:performance_comparison_2}. 
Among the models, \textbf{HICG-none} directly applies the output results from each sub-graph network as the user current interests and long-term preferences. 
\textbf{HICG-intra} only applies the intra-attention while \textbf{HICG-inter} keeps the inter-attention module. 

\begin{table}[ht]
    \centering
    \setlength{\tabcolsep}{4.9mm}
    \renewcommand\arraystretch{1.3}
    \begin{tabular}[3\columnwidth]{ccc}
        \hline
          \textbf{Yoochoose}  
          &\textbf{HR@20} &\textbf{MRR@20} \\
        \hline
        HICG-none & 62.51 & 24.04  \\
        HICG-intra & 70.79 & 29.42  \\
        HICG-inter & 68.95 & 27.65  \\
        \textbf{HICG} & \textbf{73.19} & \textbf{31.27}  \\
        \hline
    \end{tabular}

    \vspace*{2em}
    \begin{tabular}[3\columnwidth]{ccc}
        \hline
          \textbf{Diginetica}  
          &\textbf{HR@20} &\textbf{MRR@20} \\
        \hline
        HICG-none & 62.51 & 18.56  \\
        HICG-intra & 51.13 & 20.24  \\
        HICG-inter & 50.24 & 19.77  \\
        \textbf{HICG} & \textbf{51.21} & \textbf{20.51}  \\
        \hline
    \end{tabular}
  \caption{\label{tab:performance_comparison_2} Ablation Study of heterogeneous information crossing.}
\end{table}

Two key conclusions can be found from the experimental results. 
Firstly, the HICG algorithm that applies the attention mechanism is generally better than not using it, which demonstrates user's current interest more related to the latest behaviors; 
Secondly, HICG-inter has a more significant recommendation effect than HICG-none and HICG-intra by applying inter-attention mechanisms between behavior sequences which indicates there is a correlation between the user's different types of behaviors.

%% file: content/4_experiments/ra_3.tex
\begin{figure}[ht]
    \centering
    \includegraphics[width=0.48\linewidth]{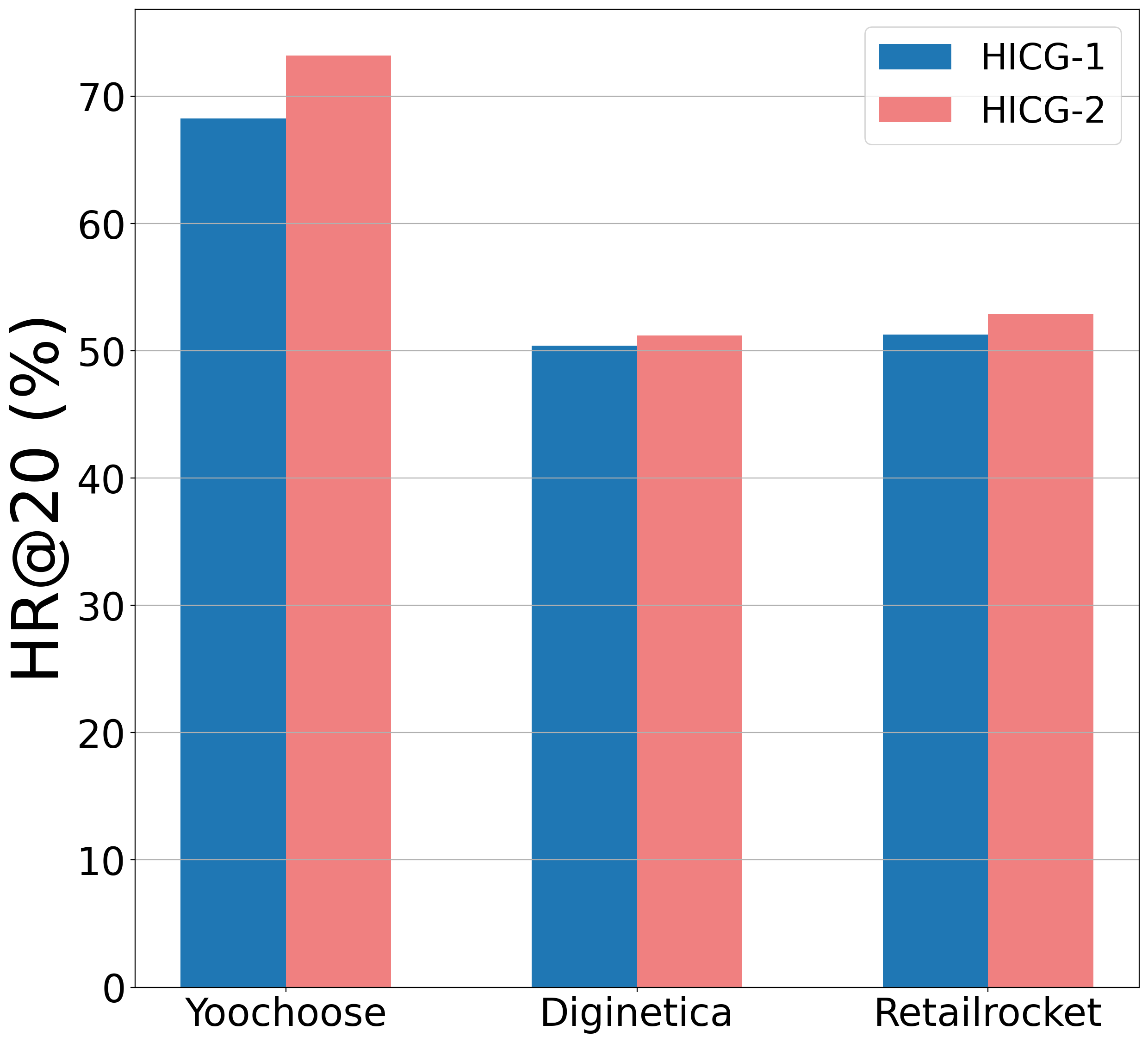}
    \includegraphics[width=0.48\linewidth]{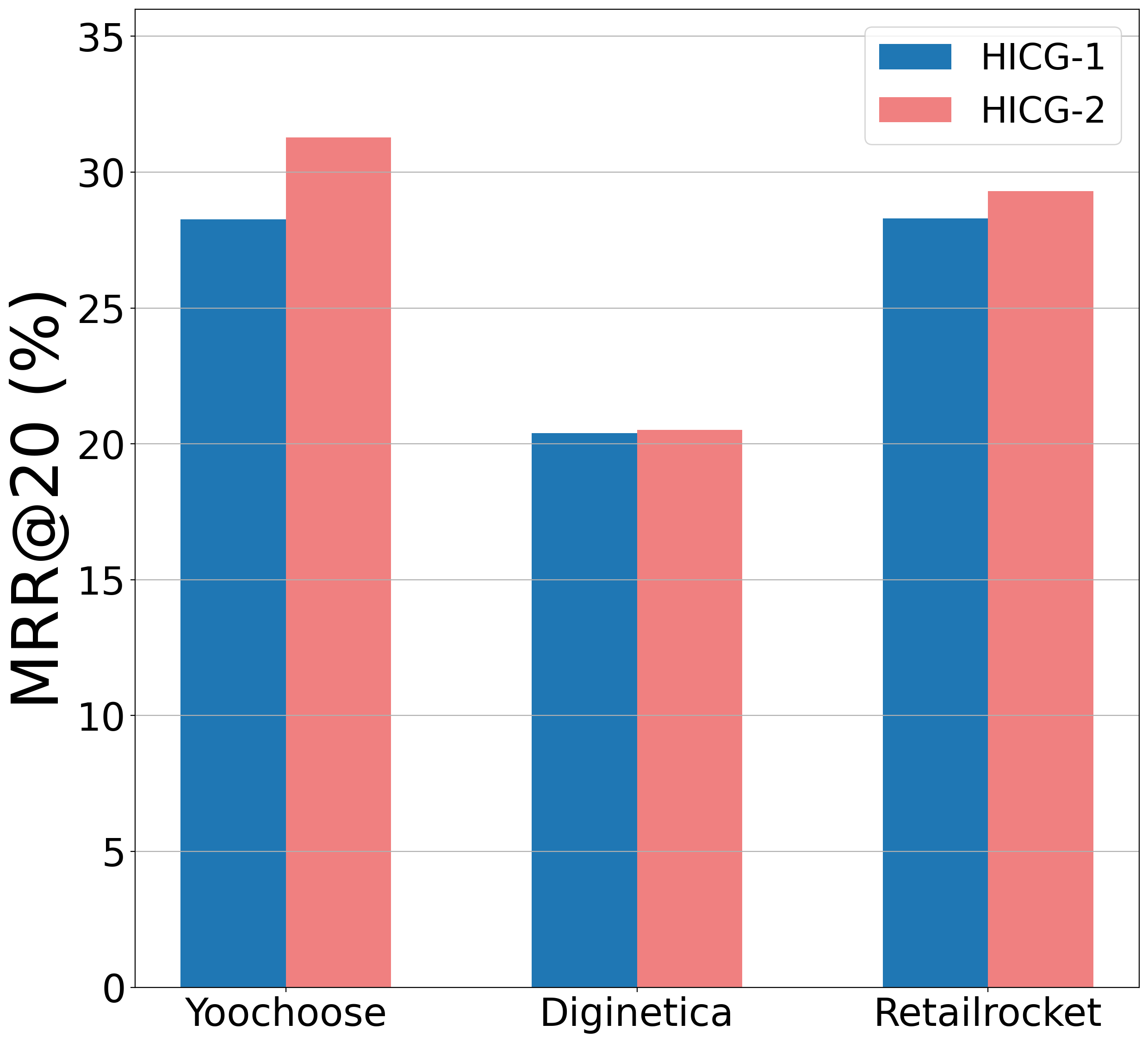}
    \caption{\label{fig:fig_3_3_1_a_3} The effect of different heterogeneous graph construction approaches on HR@20 and MRR@20.}
\end{figure}

\subsubsection{The influence of different heterogeneous graph construction approaches (RQ3)}
We conduct this experiment to analyze the effectiveness of the two proposed heterogeneous graph construction approaches, i.e., (1) ``extend the graph without crossing behavior types'' and (2) ``extend the graph with crossing behavior types''.
Briefly speaking, approach (1) constructs edges independently for each type of behavior, whereas approach (2) constructs edges using multiple types of behaviors.
We refer to the models that use these two approaches to construct the heterogeneous graph as \textbf{HICG-1} and \textbf{HICG-2}, respectively.

%
%

We report the results in \autoref{fig:fig_3_3_1_a_3}.
The results show that HICG-2 outperforms HICG-1 consistently on all datasets, which demonstrates the effectiveness of the edge-constructing method in approach 2.
The results also show that HICG-1 achieves comparable performance with HICG-2 when the number of sessions in the dataset is small (such as Diginetica). 
By comparing the statistics results of these evaluation datasets in \autoref{tab:datasets_stat}, it can be found that there are few other (purchasing) behavior data in the Diginetica dataset, and in this case, both of the construction methods have very sparse edges in the corresponding graph structure.
Therefore, the characterization ability and effect improvement brought by the introduction of heterogeneous signals are limited. 
%

%% file: content/4_experiments/ra_4.tex
\subsubsection{Performance of HICG facing different length of behavior sequences (RQ4)}
As the length of the session increases, the user may transfer interest points multiple times, making it difficult for the algorithm to understand the impact of the user's interaction on the current state of interest. 
Therefore, it is necessary to evaluate the effectiveness and robustness of the HICG algorithm for sessions of different lengths.
Applying the learned HICG model in RQ1, the effect is evaluated on the test set according to different session lengths to form subsets.

\begin{figure}[htbp]
    \centering
    \includegraphics[width=0.465\linewidth]{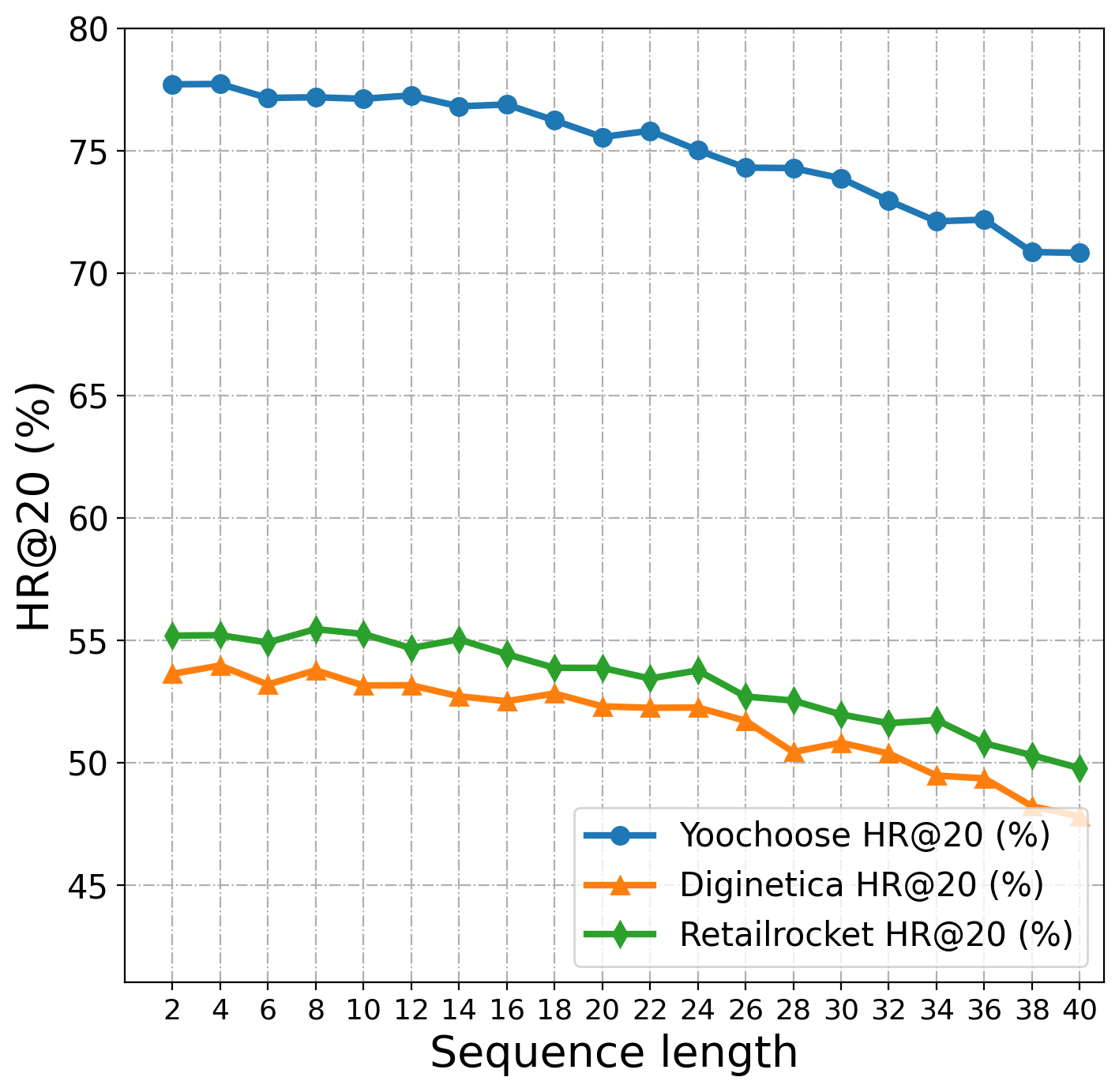}
    \includegraphics[width=0.479\linewidth]{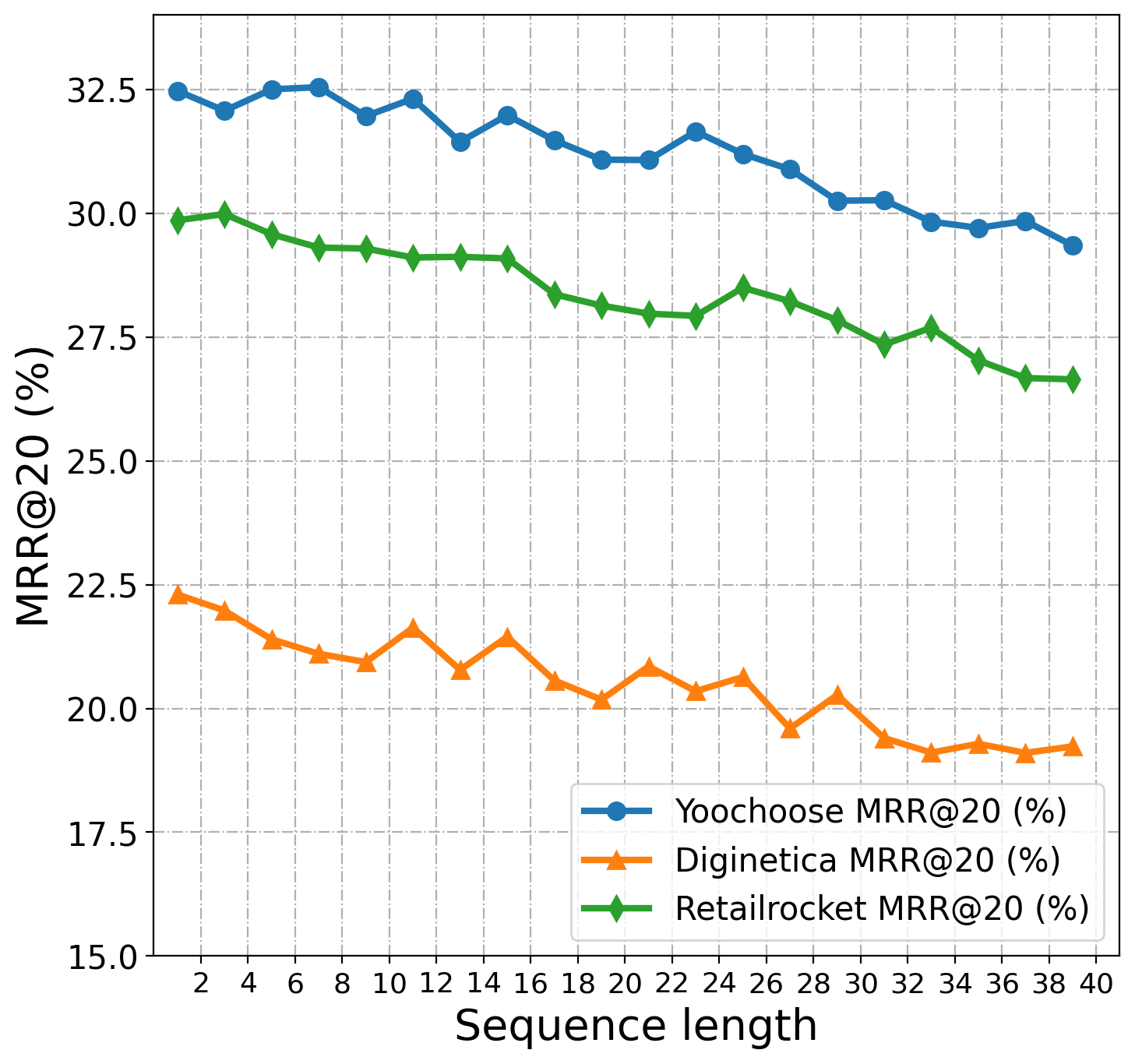}
    \caption{\label{fig:fig_3_3_3_a_6} HR@20 and MRR@20 of HICG with different sequence lengths in sessions.}
\end{figure}

The results of HR@20 and MRR@20 are shown in \autoref{fig:fig_3_3_3_a_6}.
From the experimental results, it can be seen that, in the sessions with small length of behavior sequence, HICG achieves an outstanding recommendation performance.
However, as the length of behavior sequence in sessions gradually increases, the performance of HICG gradually decreases.
This is because 
the sessions with a longer length tend to contain more different types of behaviors.
For example, there are almost only view-type behaviors in those sessions whose length is less than 5, while about 70\% of the sessions have purchase behaviors when the length is greater than 15.
As mentioned in \autoref{sss:1}, the purchase behaviors indicate that the user's corresponding need has been satisfied. 
After the purchase behavior, the subsequent change of user interests makes it more difficult to make predictions.
%

%% file: content/4_experiments/ra_6.tex
\subsubsection{The influence of the contrastive learning task (RQ5)}
In this experiment, we analyze the influence of the contrastive learning (CL) task by varying the task's weight $\lambda_{CL}$ and the sample ratio $\beta$. 
The experimental results are reported in \autoref{tab:performance_comparison_6}.
From it, we have the following two key conclusions. 
(1) The correlated CL task can lift the performance of HICG-CL when the choice of $\lambda_{CL}$ is proper (e.g., 0.1). 
However, a greater weight may instruct the model to care less about the loss in the task of next item prediction, which damages the recommendation performance.
(2) We also notice that when $\lambda_{CL} = 0.1$, a larger $\beta$ has better the recommendation performance. 
This is potentially because when the number of negative samples is larger, the NCELoss is more robust to the noise distribution ~\cite{Michael2012nce}. 
Nevertheless, we suggest to set $\beta$ to 0.1 in practice, as a larger value of it will naturally cost more time.

\begin{table}[ht]
  \centering
  \setlength{\tabcolsep}{4.9mm}
  \renewcommand\arraystretch{1.3}
  \begin{tabular}[3\columnwidth]{cccccc}
      \hline
      &  &$\lambda_{CL} = 0$ &$\lambda_{CL} = 0.1$ &$\lambda_{CL} = 0.2$ &$\lambda_{CL} = 0.3$ \\
      \hline
      & $\beta$ = 1\% & \underline{73.06} & 73.02 & 70.46 & 66.13   \\
      HR@20& $\beta$ = 10\% & 73.06 & \underline{73.19} & 70.98 & 63.04   \\
      & $\beta$ = 100\% & 73.06 & \textbf{73.28} & 70.51 & 60.71   \\
      \midrule
      & $\beta$ = 1\% & \underline{31.59} & 31.13 & 31.25 & 29.94   \\
      MRR@20& $\beta$ = 10\% & \underline{31.59} & 31.27 & 31.36 & 28.82   \\
      & $\beta$ = 100\% & 31.59 & \textbf{32.04} & 30.28 & 28.63   \\
      \hline
  \end{tabular}
  \caption{\label{tab:performance_comparison_6} Study of the influence of the contrastive learning task on Yoochoose. 
    %
    %
    The best performance is shown in bold font, while the runner-up is shown in underline. 
  }
  \end{table}

%% file: content/4_experiments/ra_5.tex
\begin{figure}[htbp]
    \centering
    \includegraphics[width=0.47\linewidth]{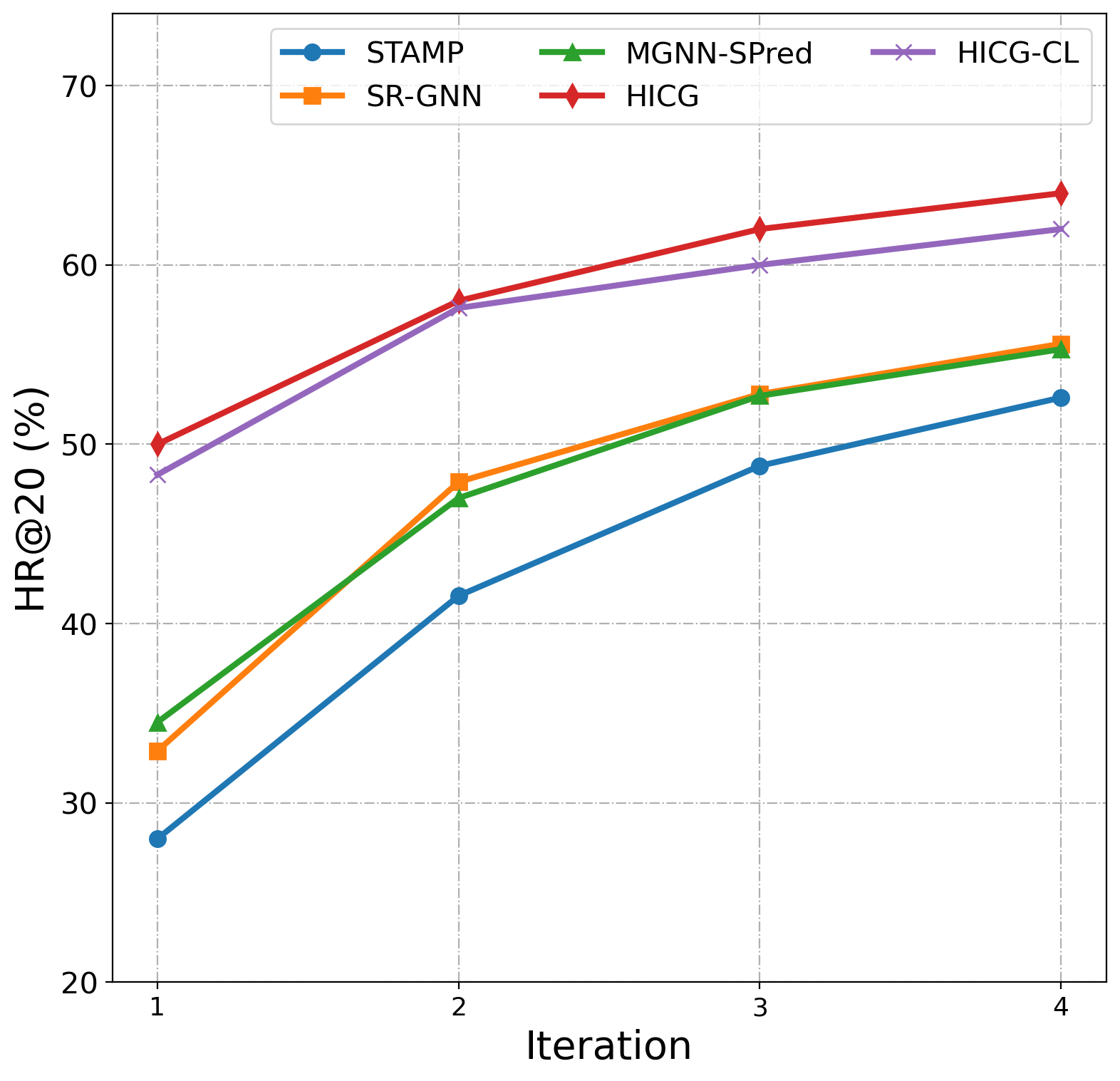}\quad \quad
    \centering
    \includegraphics[width=0.47\linewidth]{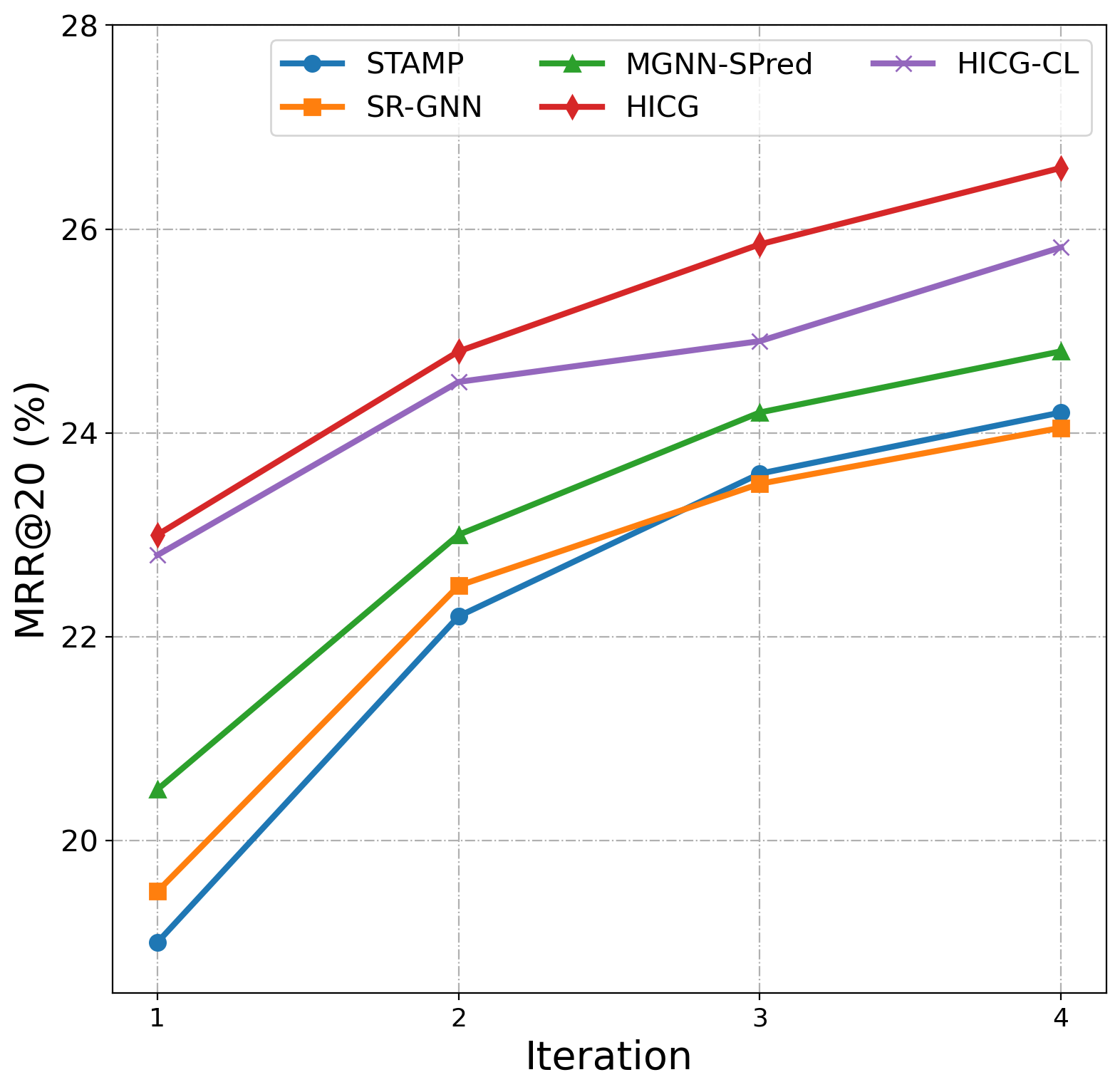}

    (a) Yoochoose
    \vspace*{1em}
    
    \centering
    \includegraphics[width=0.47\linewidth]{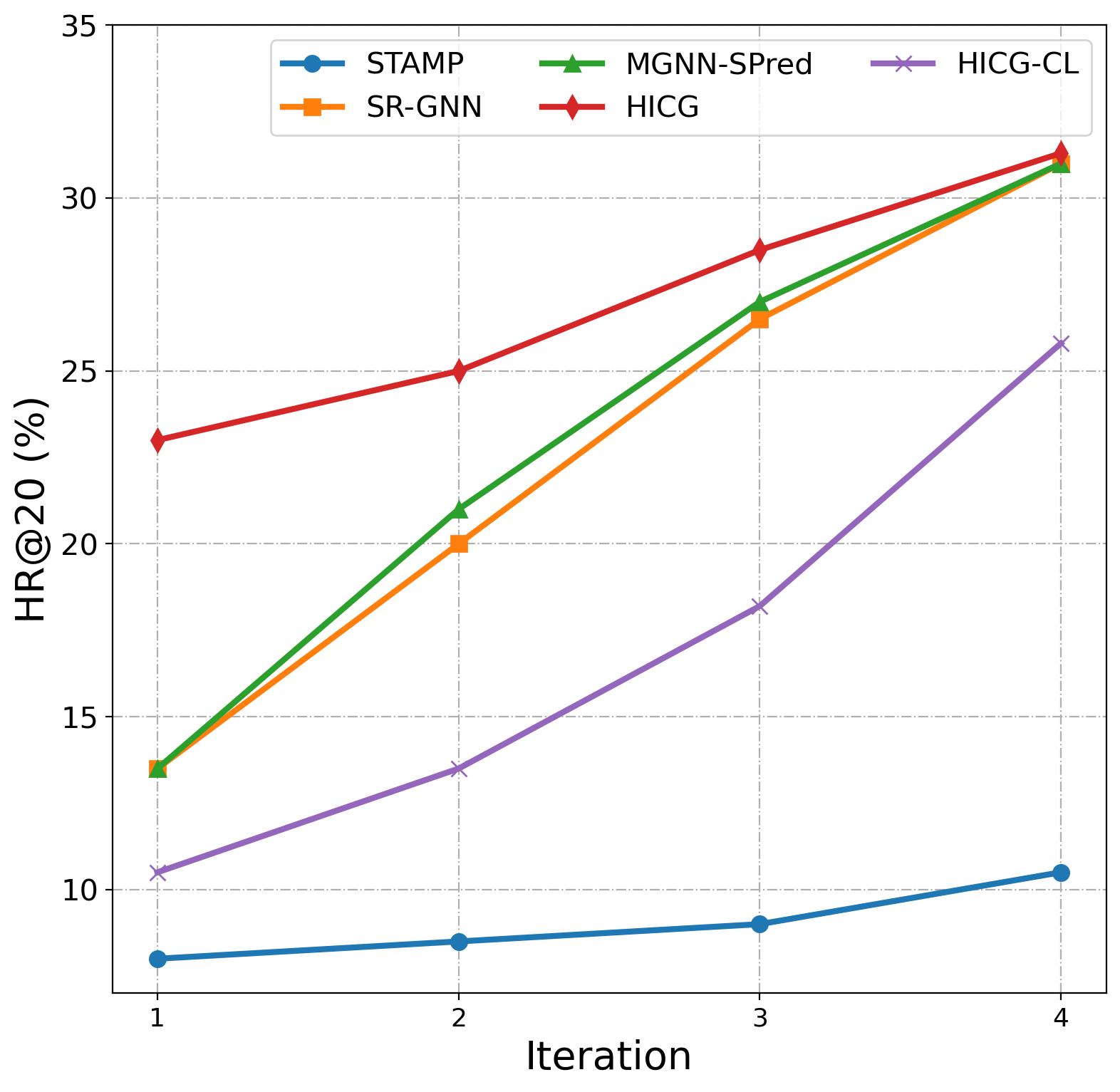}\quad \quad
    \centering
    \includegraphics[width=0.47\linewidth]{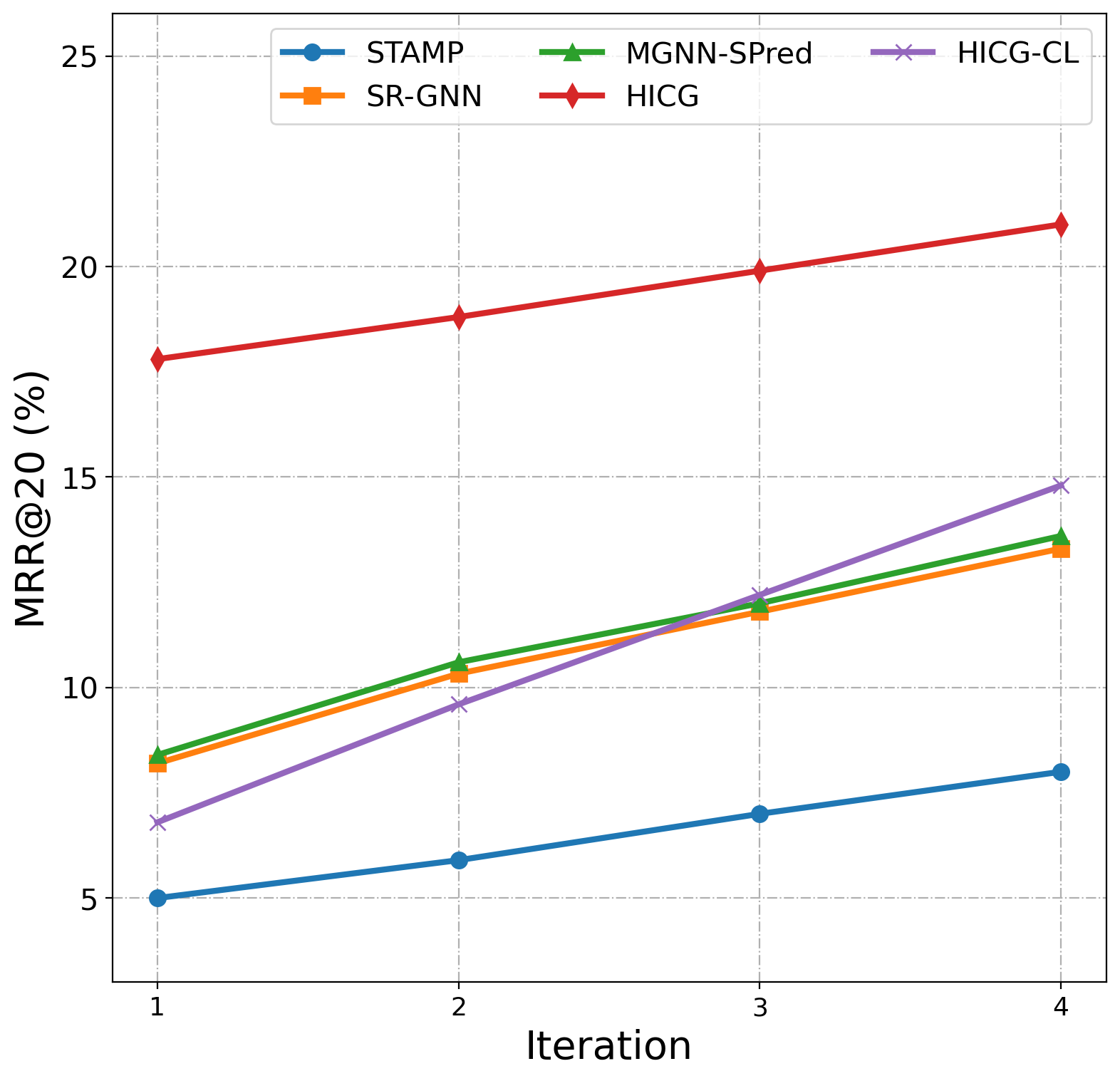}

    (b) Diginetica

    \caption{\label{fig:converge} The recommendation performance when algorithms are trained by limited iterations.}
\end{figure}

\subsubsection{The trainability of HICG algorithm (RQ6)}
We compare the recommendation performance of these algorithms that are trained by limited iterations. Our purpose is to determine whether introducing heterogeneous information in HICG and utilizing contrastive learning in HICG-CL leads to more iteration requirements, which finally affect the trainability of those methods.
The result is shown in \autoref{fig:converge}.
According to the results, we can find that both HICG and MGNN-SPred have better performance compared with other methods that employ a single type of behavior, even at the first iteration, which indicates heterogeneous information can help models fit.
Besides, in both datasets, HICG-CL starts with lower performance than our GNN-based methods but has a more significant performance lifting trend.

%% file: content/4_experiments/ra_7.tex
\subsubsection{The practicality of HICG algorithm (RQ7)}

We also conduct experimental analysis to answer RQ7.
Compared with other state-of-the-art algorithms, the HICG algorithm introduces more behavioral records and cross-extracts the relationship between different types of behaviors. 
Therefore, it is necessary to conduct corresponding experiments on the time and space efficiency of the HICG algorithm and analyze its practicality.
\autoref{tab:3.3.5} shows the time efficiency of HICG compared to other graph-based SRBS algorithms. 

\begin{table}[htbp]
    \centering
    \renewcommand\arraystretch{1.3}
    \setlength{\tabcolsep}{3mm}
    \begin{tabular}
        {crrrrrr}
        \hline \vspace*{-0.3em}
          & \multicolumn{2}{c}{\textbf{Yoochoose}} 
          & \multicolumn{2}{c}{\textbf{Diginetica}} 
          & \multicolumn{2}{c}{\textbf{Retailrocket}} \\
          \vspace*{0.3em}
          \textbf{Algorithm}
          &\multicolumn{2}{c}{\rule{0.15\columnwidth}{0.1pt}} 
          &\multicolumn{2}{c}{\rule{0.15\columnwidth}{0.1pt}} 
          &\multicolumn{2}{c}{\rule{0.15\columnwidth}{0.1pt}} \\ 
          &\textbf{Train} &\textbf{Test} &\textbf{Train} &\textbf{Test} &\textbf{Train} &\textbf{Test}\\
          \hline
        SR-GNN & 580 & 15 & 1,672 & 57 & 1,025 & 43 \\
        MGNN-SPred & 2,059 & 62 & 3,162 & 238 & 3,457 & 127 \\
        MCPRN &  1,044 & 26 & 2,364 & 81 & 2,505 & 65 \\
        HICG  & 679 & 32 & 1,758 & 83 & 1,919 & 71 \\
        HICG-CL ($\beta$ = 10\%) & 853 & 31 & 1,925 & 83 & 2,841 & 71 \\
        HICG-CL ($\beta$ = 100\%) & 3,158 & 32 & 4,390 & 84 & 5,582 & 72 \\
        \hline
    \end{tabular}
    \caption{\label{tab:3.3.5} Time costs of HICG (s).}
\end{table}

The experimental results show that compared with SR-GNN, HICG introduces a variety of behavioral data, which increases the time-consuming of the model in training and prediction. 
In addition, MGNN-SPred uses the whole relationship of items across different sessions, which is the most time-consuming in model training.
Compared with MGNN-SPred, HICG constructs heterogeneous graphs for each session, and only costs 1/3 training time while the performance is improved a lot.

\begin{table}[htbp]
    \centering
    \renewcommand\arraystretch{1.3}
    \setlength{\tabcolsep}{4.3mm}
    \begin{tabular}
        {cccc}
        \hline
        \textbf{Algorithm}  &\textbf{Yoochoose} &\textbf{Diginetica} &\textbf{Retailrocket}\\
        \hline
        SR-GNN & 2.39 & 3.59 & 2.47 \\
        MGNN-SPred & 4.36 & 6.78 & 5.22 \\
        MCPRN &  2.75 & 4.12 & 2.97 \\
        HICG & 2.42 & 3.70 & 2.65\\
        HICG-CL ($\beta$ = 10\%)  & 2.59 & 3.78 & 2.83\\
        HICG-CL ($\beta$ = 100\%)  & 3.62 & 4.91 & 3.17\\
        \hline
    \end{tabular}

    \caption{\label{tab:3.3.6} GPU memory costs of HICG (GB).}

\end{table}

In addition, \autoref{tab:3.3.6} records the GPU memory usage of HICG compared to other graph model algorithms during training.
It can be seen from the experimental results that the HICG algorithm needs to calculate heterogeneous information compared with SR-GNN,
so the GPU memory occupies a relatively large amount, but its requirements are still lower than other graph-based or heterogeneous behaviors utilizing algorithms.

%% file: content/5_conclusion.tex
\section{Conclusion} \label{ss:5}
We propose a novel heterogeneous graph-based method named HICG for SRBS.
By constructing heterogeneous graphs and applying a heterogeneous information crossing approach, HICG utilizes different types of user behaviors and learns the relationships between them. 
We also propose an enhanced version, named HICG-CL, which employs a contrastive learning technique to enhance the learning of item representations.
We finally conduct extensive experiments, on which the state-of-the-art performance achieved demonstrates the effectiveness and superiority of the proposed method.

Our article is a preliminary attempt to utilize different types of user behaviors by heterogeneous graph modeling for session-based recommendations. 
Several directions for future research are promising: 
First, the proposed algorithm HICG-CL exploits item cross-session information by constructing a joint session graph during the training phase, which means it still utilizes the information at a session level.
The question of how to apply this item cross-session information directly at a modeling level to enhance the recommendation effect is worthy of further research.
Second, when applied in the practical applications, SBRS requires online calculating of user preferences based on recent behaviors, which is time-consuming. 
Therefore, how to combine SBRS with knowledge distillation and other methods to effectively extract a lighter model is a direction with great application value.

%% file: content/6_acknowledgments.tex
\begin{acks}
This work was supported in part by the National Natural Science Foundation of China (No. 62172362), the Leading Expert of ``Ten Thousands Talent Program'' of Zhejiang Province (No.2021R52001), and was supported in part by the cooperation project of MYbank, Ant Group.
\end{acks}